\documentclass[aps,prd,twocolumn,preprintnumbers,nofootinbib,superscriptaddress,amsmath]{revtex4-2}

\usepackage{graphicx}
\usepackage{amsmath}
\usepackage{amssymb}
\usepackage{amsfonts}
\usepackage[colorlinks]{hyperref}
\usepackage{url}
\usepackage[dvipsnames]{xcolor}
\usepackage{bm}
\usepackage{array}
\usepackage{placeins}
\usepackage{bbold}
\usepackage{algorithm}
\usepackage{algpseudocode}
\usepackage[normalem]{ulem}
\usepackage{xspace}
\usepackage{orcidlink}
\usepackage{ragged2e}

\definecolor{dodgerblue}{HTML}{1E90FF}
\definecolor{crimson}{HTML}{DC143C}
\definecolor{auwien}{HTML}{8B64DD}

\hypersetup{
    colorlinks=true,
    linkcolor=auwien,
    filecolor=auwien,
    citecolor = auwien,      
    urlcolor=auwien,
}

\newcommand{\ssout}[1]{}

\def\d{\mathrm{d}}
\newcommand{\rmi}{\mathrm{i}}
\newcommand{\rme}{\operatorname{e}}
\allowdisplaybreaks[4]

\newcommand{\ord}[1]{\mathcal{O} \left( #1 \right)}

\newcommand{\av}[1]{\left\langle #1 \right\rangle}

\begin{document}
\makeatother

\preprint{IFT-UAM/CSIC-25-68}

\title{Modeling Gravitational Wave Modes from the Inspiral of Binaries with Arbitrary Eccentricity}

\author{Gonzalo Morras \orcidlink{0000-0002-9977-8546}}
\email{gonzalo.morras@uam.es}
\affiliation{Instituto de F\'isica Te\'orica UAM/CSIC, Universidad Aut\'onoma de Madrid, Cantoblanco 28049 Madrid, Spain}

\date{\today}

\begin{abstract}

Eccentric binaries are key targets for current and future gravitational wave (GW) detectors, offering unique insights into the formation and environments of compact binaries. However, accurately and efficiently modeling eccentric waveforms remains challenging, in part due to their complex harmonic structure. In this work, we develop a post-Newtonian (PN) framework to compute the Fourier amplitudes of GWs from the inspiral of eccentric binaries, deriving simple expressions at 1PN order for all relevant $(l, m)$ multipoles, valid for arbitrary eccentricities. We then characterize the GW emission by analyzing the contribution of each $(l, m)$ mode to the strain, its mean frequency, frequency spread, and asymptotic behavior at high frequencies. Additionally, we introduce a method to determine the minimal set of Fourier modes needed to reconstruct the waveform to a given accuracy. Finally, we discuss how our framework can be extended to higher PN orders, obtaining closed-form expressions for the leading-order tail and spin contributions and outlining the steps required to include higher-order corrections. Our results provide both a deeper theoretical understanding of eccentric GW emission and practical tools for developing more accurate and efficient waveform models.

\end{abstract}
\maketitle

\section{Introduction}
\label{sec:intro}

The detection of GWs from compact binary coalescences (CBCs) by the LIGO-Virgo-KAGRA (LVK) collaboration~\cite{LIGOScientific:2014pky,VIRGO:2014yos,KAGRA:2020tym} has ushered in a new era in astrophysics and fundamental physics~\cite{LIGOScientific:2016aoc,LIGOScientific:2017vwq}. Among the more than one hundred events observed to date~\cite{LIGOScientific:2018mvr,LIGOScientific:2020ibl,KAGRA:2021vkt}, most are consistent with quasi-circular inspirals, due to the circularizing effect of GW emission over time~\cite{Peters:1963ux,Peters:1964zz}. Nonetheless, there is growing observational evidence that some systems retain non-negligible orbital eccentricity by the time they enter the sensitive band of LVK detectors~\cite{Gamba:2021gap,Romero-Shaw:2020thy,Gayathri:2020coq,Romero-Shaw:2022xko,Gupte:2024jfe,Morras:2025xfu,Planas:2025jny,Planas:2025plq}. Moreover, future detectors with improved low-frequency sensitivity, such as Cosmic Explorer~\cite{Evans:2021gyd}, Einstein Telescope~\cite{Abac:2025saz}, or LISA~\cite{LISA:2024hlh}, will be able to observe binaries earlier in their inspiral, before gravitational radiation has had time to circularize their orbits. As a result, they are expected to detect systems with significantly higher orbital eccentricities~\cite{LISA:2022yao,Xuan:2023azh,Saini:2023wdk}.

Modeling orbital eccentricity is a key priority in the GW community, as it provides a relatively clean signature of the astrophysical formation channels and environments of compact binaries~\cite{Naoz:2016klo,Samsing:2017xmd,Tagawa:2020jnc,Zevin:2021rtf}. Furthermore, neglecting eccentricity in waveform models can introduce significant biases in GW searches~\cite{Divyajyoti:2023rht,Phukon:2024amh,Gadre:2024ndy}, parameter estimation~\cite{Favata:2021vhw}, and precision tests of General Relativity~\cite{Saini:2022igm,Narayan:2023vhm,Shaikh:2024wyn,Bhat:2024hyb}. Despite many recent advances in the modeling of eccentric binaries, eccentric waveform models~\cite{Moore:2016qxz,Moore:2019xkm,Tiwari:2019jtz,Paul:2024ujx,Gamboa:2024hli,Nagar:2024dzj,Nagar:2024oyk,Nagar:2024dzj,Albanesi:2025txj,Morras:2025nlp,Planas:2025feq} still lag behind their quasi-circular counterparts in both efficiency and accuracy.

In this work, we aim to address one of the key phenomena that complicates eccentric waveform modeling relative to the quasi-circular case. In quasi-circular inspirals, the orbital velocity is nearly constant, and the GW modes are quasi-monochromatic, with the frequency of each $(l,m)$ mode being equal to $m$ times the orbital frequency~\cite{Blanchet:2013haa}. In contrast, for eccentric binaries, the orbital velocity varies within each orbit, rising near periastron and falling near apastron, leading to corresponding modulations in the GW emission~\cite{Moreno-Garrido:1995sxt}. As a result, the GW signal is no longer quasi-monochromatic. However, since the system remains quasi-periodic, it can still be decomposed into a Fourier series of harmonics of the orbital frequency. The computation of the amplitudes of these Fourier modes has been studied in the literature, typically involving small eccentricity expansions~\cite{Klein:2018ybm,Tiwari:2019jtz,Klein:2021jtd,Planas:2025feq} or infinite series of Bessel functions that converge slowly for large eccentricities~\cite{Tessmer:2010sh,Tessmer:2010ii,Boetzel:2017zza,Arredondo:2024nsl}. In contrast, in Ref.~\cite{Morras:2025nlp}, we found closed-form expressions for the leading PN amplitudes of the $(2,0)$ and $(2,2)$ modes, valid for arbitrarily large eccentricities.

In this paper, we generalize and formalize the methods introduced in Ref.~\cite{Morras:2025nlp} and use them to derive simple expressions for the amplitudes at 1PN order, including all relevant $(l,m)$ higher-order modes. We also develop methods to analytically characterize the GW emission of eccentric binaries, computing the contribution of each $(l,m)$ mode to the total strain, as well as its mean frequency, frequency spread, and asymptotic behavior for large frequencies. A key application of these results is to improve the efficiency of waveform generation. Specifically, we devise a method to determine the minimal set of Fourier modes needed to accurately reconstruct the signal within a specified tolerance, enabling the construction of computationally efficient eccentric waveform models.

The remainder of this paper is organized as follows. In Sec.~\ref{sec:GWAmplitudes}, we present the general PN formalism used to describe eccentric binaries and their GW emission. In Sec.~\ref{sec:1PN}, we derive analytic expressions for the Fourier mode amplitudes at 1PN order. In Sec.~\ref{sec:Properties}, we study the properties of the GW modes while determining how to find the optimal set of Fourier modes needed to accurately represent the waveform. In Sec.~\ref{sec:future}, we discuss how the computation of the amplitudes can be extended to higher PN orders. We conclude in Sec.~\ref{sec:conclusions} with a summary of our results and a discussion of potential applications. Lengthy derivations and additional technical details are provided in the appendices.

Unless otherwise specified, we work in geometric units ($G=c=1$), use boldface to denote vectors, and assume $m_1 \geq m_2$, where $m_1$ and $m_2$ are the component masses of the binary.

\section{Gravitational Waves from Eccentric Binaries}
\label{sec:GWAmplitudes}

In this section, we review the description of eccentric binaries and their GW emission within the PN framework, which sets the stage for the analyses presented in the remainder of this paper.

\subsection{Quasi-Keplerian orbits}
\label{sec:GWAmplitudes:orbits}

To model the GW emission from eccentric compact binaries at a given PN order, it is essential to describe the orbital dynamics consistently at the same PN order. This can be achieved using the quasi-Keplerian parametrization, which generalizes the classical Keplerian solution to include relativistic corrections~\cite{Damour:1985ecc,Damour:1988mr,Schaefer:1993qk,Wex:1995qk,Memmesheimer:2004cv}. At 1PN order, the quasi-Keplerian parametrization reads~\cite{Klein:2018ybm}

\begin{subequations}
\label{eq:qKparam}
\begin{align}
    r(u) & = a (1 - e_r \cos{u}) \, , \label{eq:qKparam:r} \\
    v(u) & = 2 \arctan\left[\left(\frac{1+e_\phi}{1-e_\phi}\right)^{1/2} \tan{\frac{u}{2}}\right] \, , \label{eq:qKparam:v} \\
    \phi(u)  & = (1 + k) v(u) \, , \label{eq:qKparam:phi} \\
    \ell(u) & \equiv n (t - t_0) = u - e \sin{u} \, , \label{eq:qKparam:l}
\end{align}
\end{subequations}

\noindent where the relative separation vector is given by $\bm{x} = r(\cos{\phi},\sin{\phi},0)$, $a$ is the semi-major axis, $e$ the eccentricity, $n = 2 \pi/P$ the mean motion, with $P$ being the orbital period, and $t_0$ is a constant of integration; the auxiliary variables $u$, $v$ and $\ell$ are the eccentric, true and mean anomalies. Comparing with the Keplerian parametrization, we have introduced the periastron advance $k$ and the radial and angular eccentricities, $e_r$ and $e_\phi$. At 1PN, the constants appearing in Eq.~\eqref{eq:qKparam} are given by~\cite{Klein:2018ybm}

\begin{subequations}
\label{eq:qKvars}
\begin{align}
    a & = \frac{M}{\left( 1 - e^2 \right) y^2} \left\{ 1 + \left[ -1 + \frac{\nu}{3} + \left( 3 - \frac{\nu}{3} \right) e^2 \right] y^2  \right\} \, , \label{eq:qKvars:a} \\
    n & = \frac{\left(1 - e^2\right)^{3/2} y^3}{M} \left\{ 1 - 3 y^2 \right\} \, , \label{eq:qKvars:n} \\
    e_r^2 & = e^2 \left\{ 1 + \left(1 - e^2 \right) \left(8 - 3\nu \right) y^2 \right\} \, , \label{eq:qKvars:e_r} \\
    e_\phi^2 & = e^2 \left\{1 + \left(1 - e^2 \right) \left(8 - 2\nu \right) y^2 \right\}\, , \label{eq:qKvars:e_phi} \\
    k &= 3 y^2 \, , \label{eq:qKvars:k}
\end{align}
\end{subequations}

\noindent where we have introduced the PN parameter $y$, that is related to the norm of the Newtonian angular momentum ($L_N = \nu/y$),

\begin{equation}
    y = \frac{(M \omega)^{1/3}}{\sqrt{1 - e^2}} \, , \label{eq:y_def}
\end{equation}

\noindent with $M = m_1 + m_2$ the total mass, $\nu = m_1 m_2/M^2$ the symmetric mass ratio and $\omega$ the mean orbital frequency. In the quasi-Keplerian parametrization of Eq.~\eqref{eq:qKparam}, the orbital phase $\phi$ is not $2\pi$ periodic in the eccentric anomaly $u$ due to the effect of periastron advance $k$. To make this explicit, we separate the phase $\phi$ into the mean phase $\lambda$, that grows secularly with time, and a $2 \pi$-periodic correction $W_\phi$. At 1PN these are given by

\begin{subequations}
    \label{eq:phi_terms}
    \begin{align}
        \phi & \equiv \lambda + W_\phi \, , \label{eq:phi_terms:def} \\
        \lambda & \equiv (1 + k) \ell \, , \label{eq:phi_terms:lambda} \\
        W_\phi & = (1 + k) (v - \ell) \label{eq:phi_terms:Wphi} \, .
    \end{align}
\end{subequations}

\subsection{Fourier mode decomposition}
\label{sec:GWAmplitudes:decomposition}

To separate the angular dependence of the GW emission, we decompose the GW polarizations, $h_{+,\times}$, in terms of spin-weighted spherical harmonics~\cite{Thorne:1980ru,Kidder:2007}, i.e.,

\begin{equation}
  h_+ - \rmi h_\times = \sum_{l=2}^\infty \sum_{m=-l}^l H^{lm} ~{}_{-2} Y^{lm}(\Theta, \Phi) \, ,
  \label{eq:hlm_def}
\end{equation}

\noindent where $(\Theta, \Phi)$ are the spherical angles of the GW propagation vector in the inertial binary source frame, ${}_{-2} Y^{lm}$ are the spin-weighted spherical harmonics of spin weight $-2$, and $H^{lm}$ are the GW modes. Neglecting the effect of spin precession (which enters at 1.5PN), these modes can be written as~\cite{Mishra:2015bqa}

\begin{equation}
    H^{l m}(t) \equiv h_0 \hat{H}^{lm}(t) = h_0 \rme^{-\rmi m\phi(t)} K^{l m}[u(t)] \, ,
    \label{eq:Hlm_def}    
\end{equation}

\noindent where 

\begin{equation}
    h_0 \equiv 4\sqrt{\frac{\pi}{5}} \frac{M\nu}{d_L} (M \omega)^{2/3} \, ,
\end{equation}

\noindent with $d_L$ being the luminosity distance to the binary, and $\omega$ the mean orbital angular velocity. Neglecting again spin-precession effects, the up-down symmetry of the binary implies the modes satisfy

\begin{align}
    H^{l -m} = (-1)^l (H^{l m})^{*}  \, .
    \label{eq:Hl-m_from_Hlm}
\end{align}

In practical applications, we aim to express the modes $\hat{H}^{lm}(\ell, u(\ell))$ as a function of time. This would normally require numerically solving the transcendental Eq.~\eqref{eq:qKparam:l} to find the eccentric anomaly $u$ as a function of the mean anomaly $\ell$. However, this can be avoided by expressing the GW modes as a Fourier series in $\ell$ (i.e. in harmonics of the orbital frequency), which is also advantageous for transforming the signal into the frequency domain~\cite{Klein:2018ybm}. Following Ref.~\cite{Arredondo:2024nsl}, we write

\begin{equation}
    \hat{H}^{l m} = \rme^{-\rmi m (\lambda - \ell)} \sum_{p=-\infty}^\infty N_p^{l m} \rme^{-\rmi p \ell} \, ,
    \label{eq:Hlm_Nlmp}
\end{equation}

\noindent where we have separated the factor $\rme^{-\rmi m (\lambda - \ell)}$ since it is not $2 \pi$-periodic in $\ell$~\cite{Damour:2004bz}, and have defined the Fourier series coefficients $N_p^{l m}$, also called Fourier mode amplitudes, which can be computed as

\begin{align}
    N_{p}^{l m} & = \frac{1}{2\pi}\int_{-\pi}^\pi \left( \rme^{\rmi m (\lambda - \ell)} \hat{H}^{l m} \right) \rme^{\rmi p \ell} \d \ell \nonumber \\
    & = \frac{1}{2\pi}\int_{-\pi}^\pi F^{l m}(u) \rme^{\rmi p \ell} \d \ell \, ,
    \label{eq:Nlm_p}            
\end{align}

\noindent where, for convenience, we define

\begin{equation}
    F^{l m} = \rme^{\rmi m (\lambda - \ell)} \hat{H}^{l m} = \rme^{-\rmi m \left(\ell + W_\phi\right)} K^{l m}(u) \, .
    \label{eq:Flm_def}
\end{equation}

Our definition of $N^{l m}_p$ differs from the one in Refs.~\cite{Arredondo:2024nsl,Morras:2025nlp}, where the Fourier coefficients of Eq.~\eqref{eq:Nlm_p} would be labeled as $N_{p-m}^{l m}$. We adopt the shift $p \to p + m$ to simplify expressions and interpretation. Given that $|\lambda - \ell| = k |\ell| \approx 3 y^3 |\ell| \ll |\ell|$ in Eq.~\eqref{eq:Hlm_Nlmp}, in our convention $p$ corresponds, at leading PN order, to the ratio between the GW frequency of the mode and the orbital frequency.

Finally, given Eq.~\eqref{eq:Hlm_Nlmp}, the $\hat{H}^{l m}$ mode symmetry of Eq.~\eqref{eq:Hl-m_from_Hlm} implies that the Fourier mode amplitudes satisfy

\begin{align}
    N^{l -m}_p = (-1)^l (N^{l m}_{-p})^{*}
    \label{eq:Nl-m_from_Nlm}.
\end{align}

\section{Gravitational Wave Amplitudes at 1PN Order}
\label{sec:1PN}

In this section, we derive simple expressions for the Fourier mode amplitudes at 1PN order, which will serve as the foundation for the analyses presented in the remainder of the paper. We restrict our calculation to 1PN order, as it already captures all GW modes loud enough to be detectable by current and near-future GW detectors~\cite{Mills:2020thr}, while keeping the expressions relatively simple. Nonetheless, as will be discussed in Sec.~\ref{sec:future}, the techniques developed here can be extended to compute amplitudes at higher PN orders.

At this order, the only spin-dependent correction to the strain appears in the $(l,m) = (2,1)$ mode~\cite{Khalil:2021txt,Paul:2022xfy,Henry:2023tka}, entering at 1PN order as a term proportional to the reduced effective spin difference,

\begin{equation}
    \delta\chi = \frac{m_1 \chi_1 - m_2 \chi_2}{m_1 + m_2} \, ,
    \label{eq:dchi_deff}
\end{equation}

\noindent where $\chi_i \in [-1,1]$ are the dimensionless spins of the components projected along the orbital angular momentum. While we keep this spin-dependent term in the expressions, we set $\delta\chi = 0$ during the discussion for simplicity. Since this term is typically small, it does not affect any of the conclusions reached below.

\subsection{GW modes that contribute}
\label{sec:1PN:list}

For planar binaries, each GW mode $H^{l m}$ is determined entirely by the mass-type radiative multipole moment when $l + m$ is even, and by the current-type radiative multipole moment when $l + m$ is odd~\cite{Faye:2012we}. As a result, the leading PN order of each mode is~\cite{Mishra:2015bqa}

\begin{align}
    K^{l m} \sim 
    \begin{cases}
        \ord{y^{l-2}} &, \; \text{if $l+m$ is even} \\
        \ord{y^{l-1}} &, \; \text{if $l+m$ is odd}  \\
     \end{cases}
    \, ,
    \label{eq:Klm_PN_order}
\end{align}

\noindent and to describe the GW amplitudes at 1PN order we need to include the modes listed in Table~\ref{table:ModesByPN}.

\begin{table}[h!]
\centering
\begin{tabular}{c | c   }
PN order & Modes $(l, |m|)$ \\
\hline
\hline
0   &  $(2,0)$, $(2,2)$ \\
0.5 &  $(2,1)$, $(3,1)$, $(3,3)$ \\
1   &  $(3,0)$, $(3,2)$, $(4,0)$, $(4,2)$, $(4,4)$ \\
\hline
\end{tabular}
\caption{GW modes contributing to the waveform up to 1PN order, grouped by the PN order at which each mode first appears.}
\label{table:ModesByPN}
\end{table}

Since the first non-spinning corrections to $K^{l m}$ enter at 1PN relative order, they only need to be included for the 0PN modes ($(2,0)$ and $(2,2)$). For the $(2,1)$ mode, the leading-order spin correction appears at 0.5PN relative order, so we also include it. For the remaining modes, we use their leading-order expressions. The formulas for $K^{lm}$ used in this work are taken from Refs.~\cite{Mishra:2015bqa,Khalil:2021txt} and are explicitly written in App.~\ref{sec:appendix:Klm} using our notation.

\subsection{Fourier mode coefficients}
\label{sec:1PN:Fourier}

We now compute the Fourier mode coefficients $N_p^{l m}$ at 1PN order by using in Eq.~\eqref{eq:Nlm_p} the quasi-Keplerian parametrization of Sec.~\ref{sec:GWAmplitudes:orbits} together with the $K^{l m}$ of App.~\ref{sec:appendix:Klm}. From Eq.~\eqref{eq:phi_terms:Wphi} and Eq.~\eqref{eq:qKvars:k} we have that, at 1PN order,

\begin{equation}
    W_\phi = (1 + 3 y^2) (v - \ell) \, ,
    \label{eq:Wphi_1PN}
\end{equation}

\noindent and substituting this in Eq.~\eqref{eq:Flm_def} we find

\begin{equation}
    F^{l m} = \rme^{-\rmi m (v + 3 y^2 (v - \ell) )} K^{l m}(u) \, .
    \label{eq:Flm_1PN}    
\end{equation}

The leading exponential $\rme^{-\rmi m v}$ can be determined as $\left(\rme^{-\rmi v}\right)^m$, where $\rme^{-\rmi v}$ is computed using Eq.~\eqref{eq:qKparam:v} for $v(u)$ and basic trigonometric relations 
\begin{align}
    \rme^{-\rmi v(u)} = & \frac{\cos{u} - e_\phi - \rmi \sqrt{1-e_\phi^2}\sin{u}}{1-e_\phi\cos{u}}  \nonumber \\
    = & \frac{\cos{u} - e - \rmi \sqrt{1-e^2}\sin{u}}{1-e\cos{u}} \Bigg( 1 \nonumber\\
    & \quad - \rmi \frac{e \sqrt{1-e^2} (4- \nu) \sin{u}}{1-e \cos{u}} y^2 + \ord{y^3} \Bigg) .
    \label{eq:expiv}    
\end{align}

On the other hand, the term $\rme^{-3 \rmi m y^2 (v - \ell)}$ can just be expanded to 1PN order, leading to

\begin{align}
    \rme^{-3 \rmi m y^2 (v - \ell)} = 1 - 3 \rmi m y^2 (v- \ell) +\ord{y^3} \, .
    \label{eq:expimkvml}
\end{align}

Putting Eq.~\eqref{eq:expiv} and Eq.~\eqref{eq:expimkvml} together, we obtain 

\begin{align}
    & \rme^{-\rmi m (v + 3 y^2 (v - \ell) )} = \left( \frac{\cos{u} - e - \rmi \sqrt{1-e^2}\sin{u}}{1-e\cos{u}} \right)^m \Bigg[ 1 \nonumber\\
    & - \rmi m \left(3 (v - \ell) + \frac{e \sqrt{1-e^2} (4- \nu) \sin{u}}{1-e \cos{u}} \right) y^2 + \ord{y^3} \Bigg] ,
    \label{eq:expimvpkvml}    
\end{align}

\noindent and following Eq.~\eqref{eq:Flm_1PN}, $F^{l m}$ is given by multiplying Eq.~\eqref{eq:expimvpkvml} by the $K^{l m}$ of App.~\ref{sec:appendix:Klm}. Substituting this in Eq.~\eqref{eq:Nlm_p} we obtain complicated integrals for the Fourier mode coefficients $N_p^{l m}$. Nonetheless, similarly to how it was done in Ref.~\cite{Morras:2025nlp} to compute the 0PN Fourier mode coefficients, we can use the well known property of Fourier series coefficients 

\begin{align}
    \frac{1}{2 \pi} \int_{-\pi}^\pi & \frac{\d G}{\d \ell} \rme^{\rmi p \ell} \d\ell = \frac{-\rmi p}{2\pi} \int_{-\pi}^\pi  G(\ell) \rme^{\rmi p \ell} \d \ell \, ,
    \label{eq:dG_fourier}    
\end{align}

\noindent to simplify these integrals. The property of Eq.~\eqref{eq:dG_fourier} can be easily proven using integration by parts. Therefore, as long as we can write $F^{l m}$ as derivatives with respect to $\ell$ of functions whose Fourier series coefficients we know, we can compute $N_p^{l m}$ analytically. To find such expressions, we just appropriately integrate $F^{l m}$ with respect to $\ell$ using that

\begin{align}
    G(u) & = \int g(u(\ell)) \d\ell = \int g(u) \frac{\d\ell}{\d u} \d u \nonumber\\
    & =  \int (1 -e \cos{u}) g(u)  \d u \, ,
    \label{eq:gu_integral}    
\end{align}

\noindent where we have used the 1PN expression for $\ell(u)$ of Eq.~\eqref{eq:qKparam:l}. In App.~\ref{sec:appendix:intFLm} we show $F^{l m}$ in this simplified way, and we can observe that they can be written as sums of terms of the form $\rme^{\rmi n u}/(1 - e \cos{u})$ for $n \in \mathbb{Z}$. The Fourier series coefficients for such terms are given by

\begin{align}
    \frac{1}{2\pi}\int_{-\pi}^\pi \frac{\rme^{\rmi (n u + p \ell)}}{1 - e \cos{u}} \d \ell & = \frac{1}{2 \pi} \int_{-\pi}^\pi \rme^{\rmi [(p + n) u - p e \sin{u} ]} \d u \nonumber \\
    & = J_{p + n}(p e)
    \label{eq:enu_basic_integral}    
\end{align}

\noindent where $J_q(z)$ is the Bessel function of integer order $q$~\cite{Abramowitz_and_Stegun}. Using Eq.~\eqref{eq:dG_fourier} and Eq.~\eqref{eq:enu_basic_integral} on the formulas of App.~\ref{sec:appendix:intFLm}, we obtain the simple expressions for the Fourier mode amplitudes $N^{l m}_p$ listed in App.~\ref{sec:appendix:FourierModes}. 

\begin{figure}[h!]
\centering  
\includegraphics[width=0.5\textwidth]{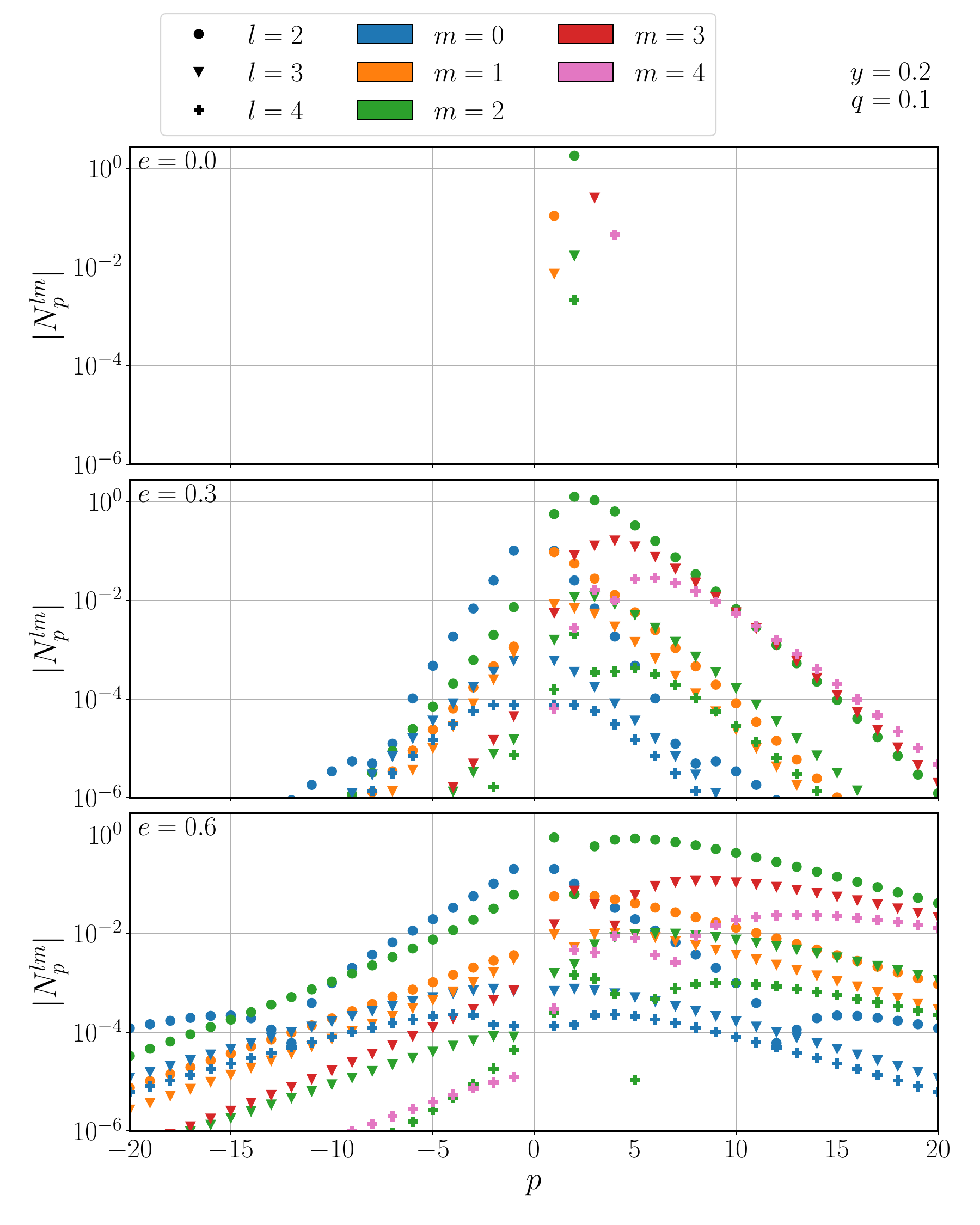}
\caption{\justifying Absolute value of the 1PN Fourier mode amplitudes, $N^{l m}_{p}$, as a function of $p$. Each panel shows $N^{l m}_{p}$ for a different value of the eccentricity $e$, with fixed PN parameter $y=0.2$ and mass ratio $q=m_2/m_1=0.1$. To compute the plotted $N^{l m}_{p}$ we have used Eq.~\eqref{eq:Np}.}
\label{fig:Nlm_p}
\end{figure}

In Fig.~\ref{fig:Nlm_p} we show the absolute value of the 1PN Fourier mode amplitudes, $N^{l m}_{p}$, as a function of $p$ and for different eccentricities. There, we can observe some of the properties of $N^{l m}_{p}$ that will be more deeply explored throughout the paper. We note that $N^{l m}_0 = 0$ for all modes, indicating the absence of a constant offset and consistent with the exclusion of GW memory~\cite{Christodoulou:1991cr,Blanchet:1992br,Favata:2011qi}. In the quasi-circular limit ($e=0$), we recover the well-known result~\cite{Blanchet:2013haa} that the GW frequency equals $m$ times the orbital frequency, i.e., $N^{l m}_p (e = 0) \propto \delta_{p m}$, with vanishing $m=0$ modes. As the eccentricity increases, the $N^{l m}_p$ amplitudes become more widely distributed in $p$, decaying exponentially as $|p| \to \infty$, but at a slower rate for larger $e$.

\section{Fourier Modes to be included}
\label{sec:Properties}

As seen in Eq.~\eqref{eq:Hlm_Nlmp}, an exact Fourier decomposition of the signal requires summing over an infinite number of modes. Since this is not feasible in practice, and to minimize the computational cost, we typically aim to include as few modes as possible while maintaining the waveform accuracy within a prescribed tolerance. This section addresses how to identify which Fourier modes should be included to achieve this goal.

\subsection{Error in the strain induced by neglecting Fourier modes}
\label{sec:Properties:WaveformErrors}

To determine which Fourier modes to include, we have to start by quantifying the error induced on the strain when selecting only a subset of them. To simplify the problem and remove angular dependencies, we study the angle-averaged squared modulus of the strain

\begin{align}
    \av{|h_+|^2 + |h_\times|^2} & \equiv  \nonumber \int_\Omega \frac{\d \Omega}{4 \pi} \left(|h_+|^2 + |h_\times|^2\right) \nonumber \\
    & = \sum_{l=2}^\infty \sum_{m=-l}^l |H^{l m}|^2 \nonumber \\
    & =  \sum_{l=2}^\infty\left(|H^{l 0}|^2 + 2\sum_{m=1}^l |H^{l m}|^2 \right) \, , \label{eq:expecth2}    
\end{align}

\noindent where the last step uses the mode symmetry in Eq.~\eqref{eq:Hl-m_from_Hlm}, while the first step uses Eq.~\eqref{eq:hlm_def} and the orthogonality of the spin-weighted spherical harmonics, i.e.

\begin{equation}
   \int \frac{\d \Omega}{4 \pi} {}_{-2}Y^{l_1 m_1} ({}_{-2}Y^{l_2 m_2})^{*} = \delta_{l_1 l_2} \delta_{m_1 m_2}  \, . \label{eq:Orthogonality_Y} 
\end{equation}

Similarly, to remove the time dependence of Eq.~\eqref{eq:expecth2}, we compute the average value of $\av{ |h_+|^2 + |h_\times|^2}$ over one orbital cycle, i.e.

\begin{align}
    \Vert \hat{h} \Vert^2 & = \frac{1}{h_0^2} \int_{-\pi}^{\pi} \frac{\d\ell}{2 \pi} \av{|h_+|^2 + |h_\times|^2}  \nonumber \\
    & = \sum_{l=2}^\infty\left(\Vert \hat{H}^{l 0}\Vert^2 + 2 \sum_{m=1}^l \Vert \hat{H}^{l m}\Vert^2 \right)  \, , \label{eq:normh}    
\end{align}

\noindent where, for simplicity, we have normalized by $h_0$ and defined
\begin{align}
    \Vert \hat{H}^{l m}\Vert^2 & = \int_{-\pi}^{\pi}  \frac{\d \ell}{2 \pi} |\hat{H}^{l m} (\ell)|^2 = \sum_{p=-\infty}^\infty |N^{l m}_p|^2 \, ,
    \label{eq:normHlm}
\end{align}

\noindent where we have substituted Eq.~\eqref{eq:Hlm_Nlmp} for the Fourier series of $\hat{H}^{l m}$. Substituting Eq.~\eqref{eq:normHlm} into Eq.~\eqref{eq:normh} we obtain

\begin{align}
    \Vert \hat{h} \Vert^2 & = 2 \sum_{l=2}^\infty \left(\sum_{p=0}^\infty |N^{l 0}_p|^2 + \sum_{m=1}^l \sum_{p=-\infty}^\infty |N^{l m}_p|^2  \right)  \label{eq:normh_Nlm}    \, ,
\end{align}

\noindent where we have used that, for the modes with $m=0$, the mode symmetry of Eq.~\eqref{eq:Nl-m_from_Nlm} implies $|N^{l 0}_p|^2 = |N^{l 0}_{-p}|^2$. Neglecting Fourier modes in Eq.~\eqref{eq:normh_Nlm} leads to a decrease of $\Vert \hat{h} \Vert^2$ with respect to the exact value that can be computed with Eq.~\eqref{eq:normh}. The relative strain error induced by including only a selected set of modes is given by

\begin{align}
    \Delta_h & \equiv \frac{\Vert \hat{h} \Vert^2 - 2 \sum_{l} \sum_{m \geq 0} \sum_{p \in \bm{p}_{l m}^\mathrm{sel}} |N^{l m}_p|^2}{\Vert \hat{h} \Vert^2}
    \label{eq:strain_error_def}  \, ,
\end{align}

\noindent where $\bm{p}_{l m}^\mathrm{sel}$ denotes the subset of Fourier modes selected for each $(l,m)$ GW mode. For the $m = 0$ case, only $p \geq 0$ are considered due to the mode symmetry. Typically, the goal is to include as few Fourier modes as possible, while having $\Delta_h $ under a certain tolerance. Given the form of Eq.~\eqref{eq:strain_error_def}, this can be optimally achieved by progressively selecting the $(l,m,p)$ modes with largest $|N^{l m}_p|^2$ until $\Delta_h$ drops below the target threshold. In practice, this is nontrivial because the relevant $p$-range for each $(l,m)$ is not known a priori. This challenge will be addressed in the following subsections.

In Ref.~\cite{Morras:2025nlp}, we found that the strain error $\Delta_h$ is closely related to the error in the log-Likelihood, having 

\begin{equation}
    \Delta \log\mathcal{L}  \sim \rho_\mathrm{opt}^2 \Delta_h \, ,
    \label{eq:DlogL_Deltah}
\end{equation}

\noindent where $\rho_\mathrm{opt}$ is the optimal signal-to-noise ratio (SNR) of the signal under study. Therefore, minimizing $\Delta_h$ is not only convenient from a theoretical standpoint, due to the simplicity of Eq.~\eqref{eq:strain_error_def}, but is also well motivated for GW data analysis applications. As long as $\Delta \log\mathcal{L} \lesssim 1$, waveform differences have a negligible impact on event significance in searches and parameter estimation posteriors~\cite{Jaranowski:2005hz,Morras:2023pug}. For instance, with a strain error of $\Delta_h \sim 10^{-4}$, waveform inaccuracies remain negligible for signals with SNRs up to 100.

\subsection{Norms and frequency structure of GW modes}
\label{sec:Properties:Properties}

In order to estimate the Fourier modes that should be included, we first need to compute $\Vert  \hat{h} \Vert^2$. Using the last equality in Eq.~\eqref{eq:normHlm} to estimate $\Vert \hat{H}^{l m}\Vert^2$ would not help, as it still requires summing over an infinite number of Fourier modes. Instead, we employ the integral in the first equality of Eq.~\eqref{eq:normHlm}, which, in terms of $F^{l m}$, is given by

\begin{equation}
    \Vert \hat{H}^{l m}\Vert^2 = \int_{-\pi}^{\pi}  \frac{\d \ell}{2 \pi} |F^{l m}|^2 \, ,
    \label{eq:normHlm_Flm}
\end{equation}

\begin{figure}[h!]
\centering  
\includegraphics[width=0.5\textwidth]{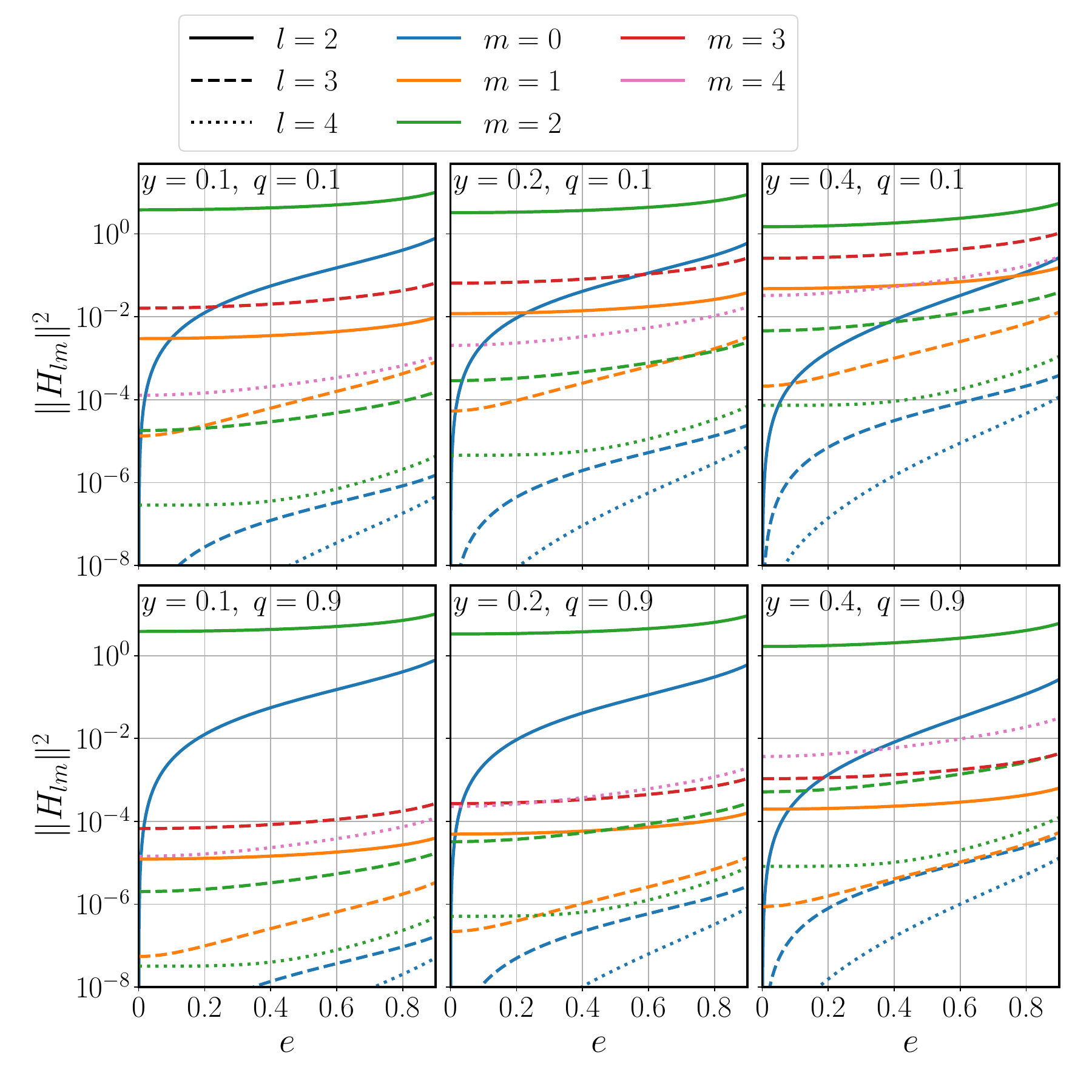}
\caption{\justifying Norm of each 1PN GW mode, $\Vert \hat{H}^{l m}\Vert^2$, as a function of eccentricity $e$. Each panel shows $\Vert \hat{H}^{l m}\Vert^2$ for specific values of the PN parameter $y$ and mass ratio $q = m_2/m_1$. To compute $\Vert \hat{H}^{l m}\Vert^2$ we have used Eq.~\eqref{eq:ezNorms}.}
\label{fig:norms_Hlm}
\end{figure}

In App.~\ref{sec:appendix:Moments:Norms} we compute the norms of the 1PN GW modes with this integral, obtaining closed form expressions. In Fig.~\ref{fig:norms_Hlm} we plot these norms as a function of eccentricity $e$ for different values of the PN parameter $y$ and the mass ratio $q=m_2/m_1$. The values of $y$ and $q$ shown in Fig.~\ref{fig:norms_Hlm} will be used throughout the rest of the paper. The $q=0.9$ case represents a nearly equal mass binary, while $q=0.1$ is around the mass ratio of GW190814~\cite{LIGOScientific:2020zkf}, representing the most extreme mass ratios that have been confidently observed to date~\cite{KAGRA:2021vkt}. Meanwhile, $y=0.4$ represents a binary close to the innermost stable circular orbit (ISCO), since $y_\mathrm{ISCO}=6^{-1/2} \approx 0.408$~\cite{Morras:2025nlp}, while $y=0.1$ represents the early inspiral and $y=0.2$ an intermediate regime. Using Eq.~\eqref{eq:y_def} we can convert $y$ to an orbital frequency, obtaining

\begin{align}
    f_\mathrm{orb} & = \frac{1}{2 \pi} \frac{c^3}{G M} (1-e^2)^{3/2} y^3 \nonumber \\
    & = 10.8 \, \mathrm{Hz} \, (1-e^2)^{3/2} \left( \frac{3 M_\odot}{M} \right) \left( \frac{y}{0.1} \right)^3 \nonumber \\
    & = 12.9 \, \mathrm{Hz} \, (1-e^2)^{3/2} \left( \frac{20 M_\odot}{M} \right) \left( \frac{y}{0.2} \right)^3 \nonumber \\
    & = 34.5 \, \mathrm{Hz} \, (1-e^2)^{3/2} \left( \frac{60 M_\odot}{M} \right) \left( \frac{y}{0.4} \right)^3 \, ,
    \label{eq:forb_of_y}
\end{align}

\noindent and therefore $y=0.1$ corresponds to a typical binary neutron star with $m_1 = m_2 = 1.5 M_\odot$ and $y=0.2$ corresponds to a typical low mass binary black hole (BBH) with $m_1 = m_2 = 10 M_\odot$ as they enter the LIGO band, while $y=0.4$ corresponds to a typical $m_1 = m_2 = 30 M_\odot$ BBH in the most sensitive part of the LIGO band.

In Fig.~\ref{fig:norms_Hlm} we observe that the norms generally increase with eccentricity, with $m=0$ modes having zero norm at $e=0$, and the norm of all modes diverging like $(1 - e^2)^{-1/2}$ as $e \to 1$, as can be seen from Eq.~\ref{eq:ezNorms}. Furthermore, as expected, the $(l,m)=(2,2)$ mode dominates across parameter space, with the $(3,3)$, $(2,1)$ and $(4,4)$ modes becoming more significant at small mass ratios and high PN parameter values. At large eccentricity, the $(2,0)$ has a very significant contribution, irrespective of the value of $y$ and $q$, since it enters at Newtonian order.

To estimate which values of $p$ contribute the most for each $(l,m)$ GW mode, we note that

\begin{equation}
    f_p^{l m} = \frac{|N^{l m}_p|^2}{\Vert \hat{H}^{l m}\Vert^2} \, \quad (p \in \mathbb{Z}) \, , 
    \label{eq:p_k}
\end{equation}

\noindent is always non negative and its sum over $p \in \mathbb{Z}$ is equal to 1. Therefore, $f_p^{l m}$ can be interpreted as a discrete probability mass function (PMF), describing how the amplitude of each GW mode is distributed in $p$. We can study this distribution by looking at its mean $\mu_{l m}$ and standard deviation $\sigma_{l m}$, given by

\begin{subequations}
\label{eq:mus_M}
\begin{align}
    \mu_{l m} & = M^{l m}_1 / M^{l m}_0 \, , \label{eq:mus_M:mu} \\
    \sigma_{l m} & = \sqrt{M^{l m}_2/M^{l m}_0 - (\mu_{l m})^2}  \, , \label{eq:mus_M:s}    
\end{align}
\end{subequations}

\noindent which provide a measure of the average and spread of the ratio between the GW and orbital frequencies for each mode. In Eq.~\eqref{eq:mus_M} we have introduced the unnormalized moments of the $f_p^{l m}$,

\begin{equation}
    M^{l m}_n = \sum_{p=-\infty}^\infty p^n |N^{l m}_p|^2 \, , 
    \label{eq:Mlm_n_def}
\end{equation}

\noindent with $M^{l m}_0 = \Vert \hat{H}^{l m}\Vert^2$. Using the mode symmetry formula of Eq.~\eqref{eq:Nl-m_from_Nlm} in Eq.~\eqref{eq:Mlm_n_def}, it is easy to show that these unnormalized moments satisfy

\begin{equation}
    M^{l -m}_{n} = (-1)^{n} M^{l m}_n \, , 
    \label{eq:Ml-m_from_Mlm}
\end{equation}

To compute the unnormalized moments, we substitute Eq.~\eqref{eq:Nlm_p} for $N^{l m}_p$ in Eq.~\eqref{eq:Mlm_n_def}, obtaining

\begin{align}
    M^{l m}_n &= \sum_{p=-\infty}^\infty \left(p^{n-q} \int_{-\pi}^\pi \frac{\d \ell_1}{2\pi} F^{l m}(\ell_1) \rme^{\rmi p \ell_1} \right) \nonumber \\
    & \qquad \qquad \times \left(p^q \int_{-\pi}^\pi \frac{\d \ell_2}{2\pi} F^{l m}(\ell_2) \rme^{\rmi p \ell_2} \right)^{*} \nonumber \\
    & = \rmi^{n - 2 q} \! \! \int_{-\pi}^\pi \frac{\d \ell}{2\pi} \frac{\d^{n-q}  F^{l m}}{\d \ell^{n-q}} \!\left(\frac{\d^{q} F^{l m}}{\d \ell^{q}} \right)^{*} \! \! ,
    \label{eq:Mlm_n}
\end{align}

\noindent where $q$ is an arbitrary integer such that $0 \leq q \leq n$, and we have used Eq.~\eqref{eq:dG_fourier} to convert the factors of $p$ in derivatives, as well as the completeness of the Fourier basis
\begin{equation}
    \sum_{p=-\infty}^\infty \frac{\rme^{\rmi p (\ell_1- \ell_2)}}{2 \pi} = \delta(\ell_1 - \ell_2) \, .
    \label{eq:Fourier_completeness}
\end{equation}

When $n=0$, we can compare Eq.~\eqref{eq:Mlm_n} with Eq.~\eqref{eq:normHlm_Flm} and observe that, as expected, $M^{l m}_0 = \Vert \hat{H}^{l m}\Vert^2$. The other moments needed to determine the mean and standard deviation can be computed from Eq.~\eqref{eq:Mlm_n} as

\begin{subequations}
\label{eq:Mlm_12}
\begin{align}
    M^{l m}_1 & = \mathrm{Im}\left\{ \int_{-\pi}^\pi \frac{\d \ell}{2\pi} F^{l m} \left(\frac{\d F^{l m}}{\d \ell} \right)^{*} \right\} , \label{eq:Mlm_12:1} \\
    M^{l m}_2 & =  \int_{-\pi}^\pi \frac{\d \ell}{2\pi} \left|\frac{\d F^{l m}}{\d \ell} \right|^2 , \label{eq:Mlm_12:2}
\end{align}    
\end{subequations}

\noindent where, for simplicity, for $n=1$ we have taken the average of $q=0$ and $q=1$, while for $n=2$ we have chosen $q=1$. In App.~\ref{sec:appendix:Moments} we use these equations to compute $M^{l m}_1$ and $M^{l m}_2$ at 1PN order for the different higher order modes. We can then use these moments to compute the mean and standard deviation of $p$ with Eq.~\eqref{eq:mus_M}. 

\begin{figure}[h!]
\centering
\includegraphics[width=0.4\textwidth]{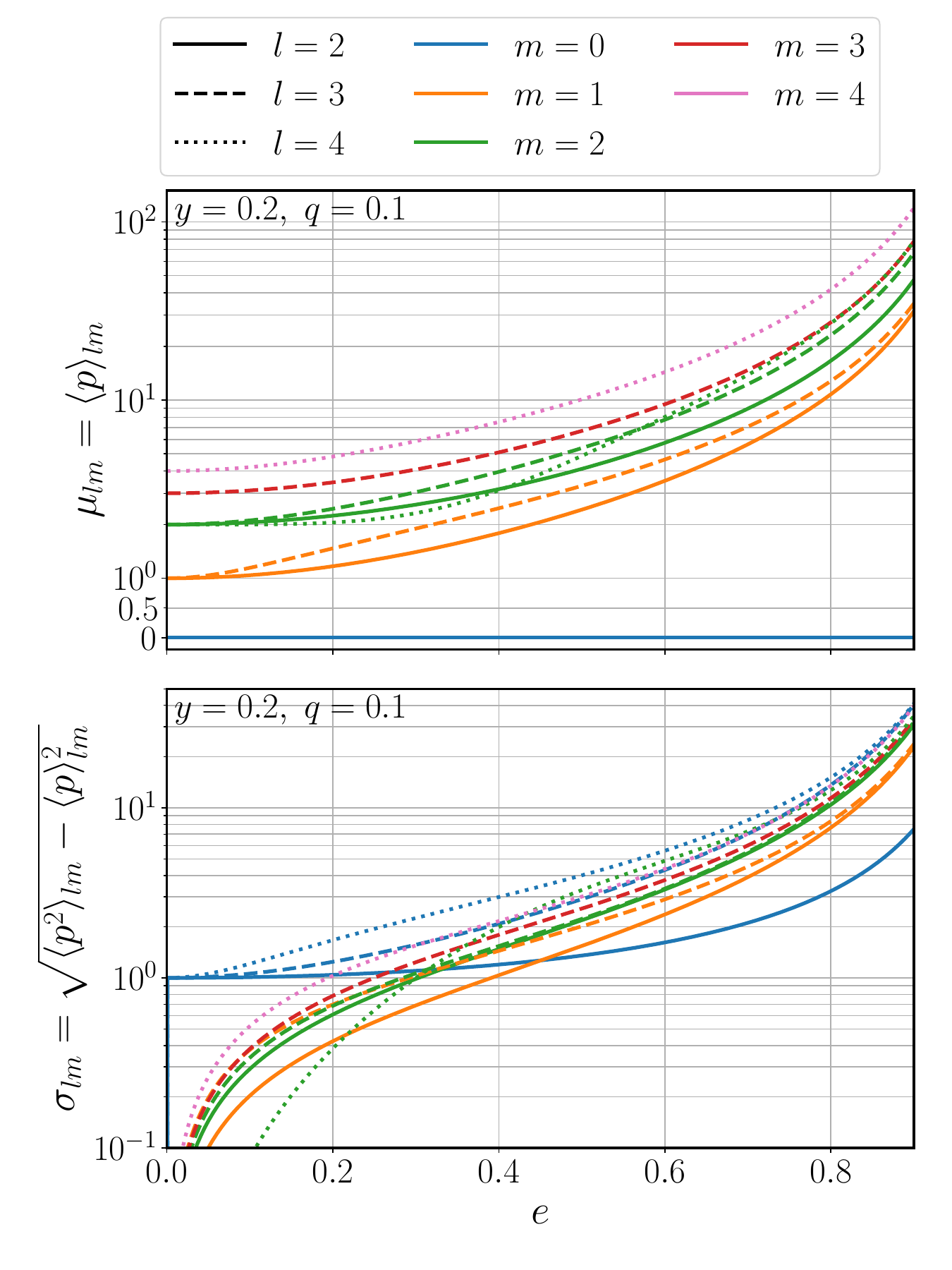}
\caption{\justifying Average, $\mu^{l m}$ (top panel), and standard deviation, $\sigma^{l m}$ (bottom panel), of $p$ for each GW mode as a function of eccentricity $e$, for a fixed value of the PN parameter ($y=0.2$) and mass ratio $(q = m_2/m_1 = 0.1)$. To compute $\mu^{l m}$ and $\sigma^{l m}$, we have used the 1PN moments of Eqs.~(\ref{eq:ezNorms},\ref{eq:ezM1},\ref{eq:ezM2}) to evaluate Eq.~\eqref{eq:mus_M}.}
\label{fig:mu_sigma_Hlm}
\end{figure}

In Fig.~\ref{fig:mu_sigma_Hlm} we show the mean $\mu^{l m}$ and standard deviation $\sigma^{l m}$ as a function of eccentricity $e$. We fix $y$ and $q$ since $\mu^{l m}$ and $\sigma^{l m}$ do not depend on these parameters at leading PN order, and, varying them would only lead to small changes in the $(2,2)$ and $(2,0)$ modes. To interpret these plots, we note that $\mu^{l m}$ measures an average ratio between the GW and orbital frequencies for each mode, while $\sigma^{l m}$ measures how spread out this ratio is.  As we saw in Fig.~\ref{fig:Nlm_p}, when $e=0$, the ratio between the GW and orbital frequencies is equal to $m$, consistent with having $\mu^{l m}(e=0) = m$ and $\sigma^{l m}(e=0) = 0$ in Fig.~\ref{fig:mu_sigma_Hlm}. Nonetheless, as the eccentricity increases, we observe that both $\mu^{l m}$ and $\sigma^{l m}$ increase, meaning that the ratio between the GW and orbital frequencies increases and becomes more spread out. This indicates that the amplitude of the GW mode comes from larger values of $p$, with more modes contributing. In particular, from the expressions in App.~\ref{sec:appendix:Moments} we can deduce that, for all GW modes, both $\mu^{l m}$ and $\sigma^{l m}$ diverge like $(1 - e^2)^{-3/2}$ as $e \to 1$.

\subsection{Conservative estimate of the required Fourier modes}
\label{sec:Properties:Estimation}

To estimate the Fourier modes needed to accurately represent the waveform, we include $n_{l m}$ of them symmetrically around the mean $\mu_{l m}$ of each GW mode. For simplicity, we temporarily ignore the mode symmetries that were used to simplify $\Delta_h$ in Eq.~\eqref{eq:strain_error_def}. This ensures all modes are treated uniformly, simplifying the analysis, but introduces a double counting, which we will correct later. With this choice, the strain error $\Delta_h$ is

\begin{align}
    \Delta_h & = \frac{\sum_{l,m} \sum_{|p - \mu_{l m}| \geq n_{l m}/2} |N^{l m}_p|^2}{\Vert  \hat{h} \Vert^2} \nonumber \\
    & = \frac{\sum_{l,m} \Vert \hat{H}^{l m}\Vert^2 \sum_{|p - \mu_{l m}| \geq n_{l m}/2} f_p^{l m}}{\Vert  \hat{h} \Vert^2} \nonumber \\
    & \leq \frac{1}{\Vert  \hat{h} \Vert^2} \sum_{l,m} \frac{4 \sigma_{l m}^2 \Vert \hat{H}^{l m}\Vert^2}{n_{l m}^2} \, , \label{eq:strain_error_Chebyshev} 
\end{align}

\noindent where the $m<0$ modes are also included in the sum, and in the last step we have used Chebyshev's inequality. Although Chebyshev's inequality usually provides rather loose bounds due to its minimal assumptions, it offers a simple conservative estimate of the required number of Fourier modes, which in this way add up to

\begin{align}
    N_F & = \sum_{l,m} 1 +  \left\lfloor \mu_{l m} + \frac{n_{l m}}{2} \right\rfloor - \left\lceil \mu_{l m} - \frac{n_{l m}}{2} \right\rceil \approx \sum_{l,m} n_{l m} .
    \label{eq:N_Fourier}    
\end{align}

\noindent where the last approximation assumes $n_{l m} \gg 1$. As previously mentioned, to reduce computational costs, we want to minimize $N_F$ while keeping the error $\Delta_h$ under a certain tolerance $\epsilon_N$, i.e.

\begin{equation}
    \Delta_h \leq \epsilon_N \, .
    \label{eq:epsilon_N_def}
\end{equation}

Using the method of Lagrange multipliers, this is equivalent to minimizing

\begin{equation}
    \mathcal{L} = \sum_{l,m} n_{lm} - \lambda \left(\epsilon_N - 4 \sum_{l,m} \frac{v_{l m}}{n_{l m}^2} \right) \, ,
    \label{eq:Lagrange_mult}
\end{equation}

\noindent with respect to $n_{l m}$ and $\lambda$, and we have defined

\begin{equation}
    v_{l m} \equiv \sigma_{l m}^2 \frac{\Vert \hat{H}^{l m}\Vert^2}{\Vert  \hat{h} \Vert^2} \, ,
\label{eq:vlm_def}
\end{equation}

\noindent as the variance of each GW mode weighted by its relative contribution to the total strain. Equating the partial derivatives of Eq.~\eqref{eq:Lagrange_mult} to zero, it is easy to show that the minimum of $\mathcal{L}$ happens when we include the following number of Fourier modes for each GW mode:

\begin{equation}
    n_{l m}^\mathrm{uncorrected} = \frac{2}{\sqrt{\epsilon_N}} v_{l m}^{1/3} \sqrt{\sum_{l' m'} v_{l' m'}^{1/3}} \, .
    \label{eq:nlm_guess_uncorr}
\end{equation}

To take into account the double counting induced by ignoring the mode symmetries, we neglect $n_{l m}$ when $m < 0$ and divide by 2 in the case of $m=0$, i.e.

\begin{align}
n_{l m}^\mathrm{guess} = \frac{1}{\sqrt{\epsilon_N}} (2 - \delta_{m,0}) v_{l m}^{1/3} \sqrt{\sum_{l', m' \geq 0} (2 - \delta_{m',0}) v_{l' m'}^{1/3}} \, .
\label{eq:nlm_guess}
\end{align}

While $n_{l m}^\mathrm{guess}$ usually overestimates the number of modes due to the looseness of Chebyshev’s bound, it provides a simple upper limit and highlights the parameters that most influence mode inclusion. Notably, a higher weighted variance $v_{l m}$ implies more modes must be included, and, due to the mode symetries, the $m=0$ modes are relatively suppressed compared to $m \geq 1$ modes.

\begin{figure}[h!]
\centering  
\includegraphics[width=0.5\textwidth]{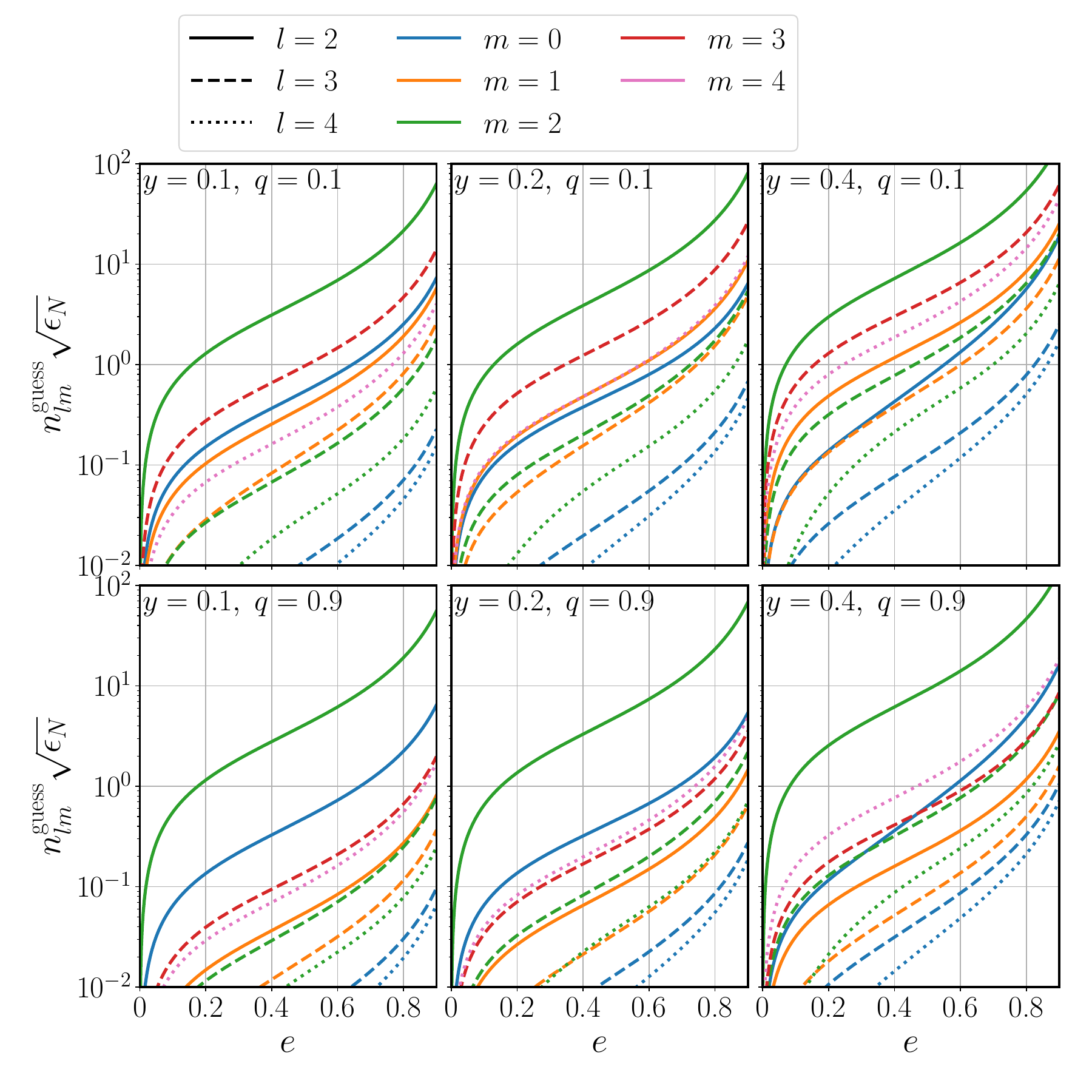}
\caption{\justifying Guess for the number of Fourier modes that have to be included for each GW mode multiplied by the square root of the tolerance, $n_{l m}^\mathrm{guess} \sqrt{\epsilon_N}$, as a function of eccentricity $e$. Each panel shows $n_{l m}^\mathrm{guess}\sqrt{\epsilon_N}$ for values of the PN parameter $y$ and mass ratio $q = m_2/m_1$ matching the configuration in Fig.~\ref{fig:norms_Hlm}. To compute $n_{l m}^\mathrm{guess} \sqrt{\epsilon_N}$, we have used the 1PN moments of Eqs.~(\ref{eq:ezNorms},\ref{eq:ezM1},\ref{eq:ezM2}) to evaluate Eq.~\eqref{eq:nlm_guess}.}
\label{fig:Nguess_Hlm}
\end{figure}

In Fig.~\ref{fig:Nguess_Hlm} we show $n_{l m}^\mathrm{guess}\sqrt{\epsilon_N}$ for the different GW modes as a function of eccentricity. We multiply $n_{l m}^\mathrm{guess}$ by $\sqrt{\epsilon_N}$ to remove the dependence on $\epsilon_N$ in Eq.~\eqref{eq:nlm_guess}. We observe that larger eccentricities require more Fourier modes, as expected from the fact that, as was seen in Fig.~\ref{fig:mu_sigma_Hlm}, the standard deviations increase with eccentricity. In particular, from the expressions in App.~\ref{sec:appendix:Moments}, we can deduce that, $n_{l m}^\mathrm{guess}$ diverges like $(1 - e^2)^{-3/2}$ as $e \to 1$, for all GW modes. Furthermore, we observe that the $(2,2)$ mode dominates the number of Fourier modes required, given that as seen in Fig.~\ref{fig:norms_Hlm}, it has by far the largest norm, and therefore has to be represented with a better relative accuracy. Nonetheless, there are many other modes for which we need a large number of Fourier modes, especially for small values of the tolerance $\epsilon_N$, large eccentricities, and large PN parameters and extreme mass ratios.

\subsection{Optimal Fourier modes to include}
\label{sec:Properties:OptimumInclusion}

In this section, we describe how to optimally select the set of Fourier modes needed to represent the strain to a given tolerance $\epsilon_N$, while minimizing the number of modes included. Since the terms in the sum of Eq.~\eqref{eq:strain_error_def} are mutually independent, this can be achieved by sequentially selecting the $(l,m,p)$ modes with the largest $|N^{l m}_p|^2$ until the residual error drops below the target tolerance, i.e.

\begin{align}
    \Delta_h & \equiv \frac{\Vert \hat{h} \Vert^2 - 2 \sum_{l} \sum_{m \geq 0} \sum_{p \in \bm{p}_{l m}^\mathrm{sel}} |N^{l m}_p|^2}{\Vert \hat{h} \Vert^2} \leq \epsilon_N
    \label{eq:strain_error_ineq}  \, .
\end{align}

To optimally select the Fourier modes in this way, the values of $|N^{l m}_p|^2$ for all necessary modes have to be tested. A possible way to guarantee this is by using the toy model developed in Sec.~\ref{sec:Properties:Estimation}, considering $n_{l m}^\mathrm{guess}$ modes around the mean $\mu_{l m}$ of each GW mode. As noted earlier, this is typically a gross overestimation. While this guarantees that the modes with largest norms are tested, ensuring that we find the optimal set, it can be computationally inefficient, as many modes with small norms are initially being considered.

\begin{figure}[h!]
\centering  
\includegraphics[width=0.5\textwidth]{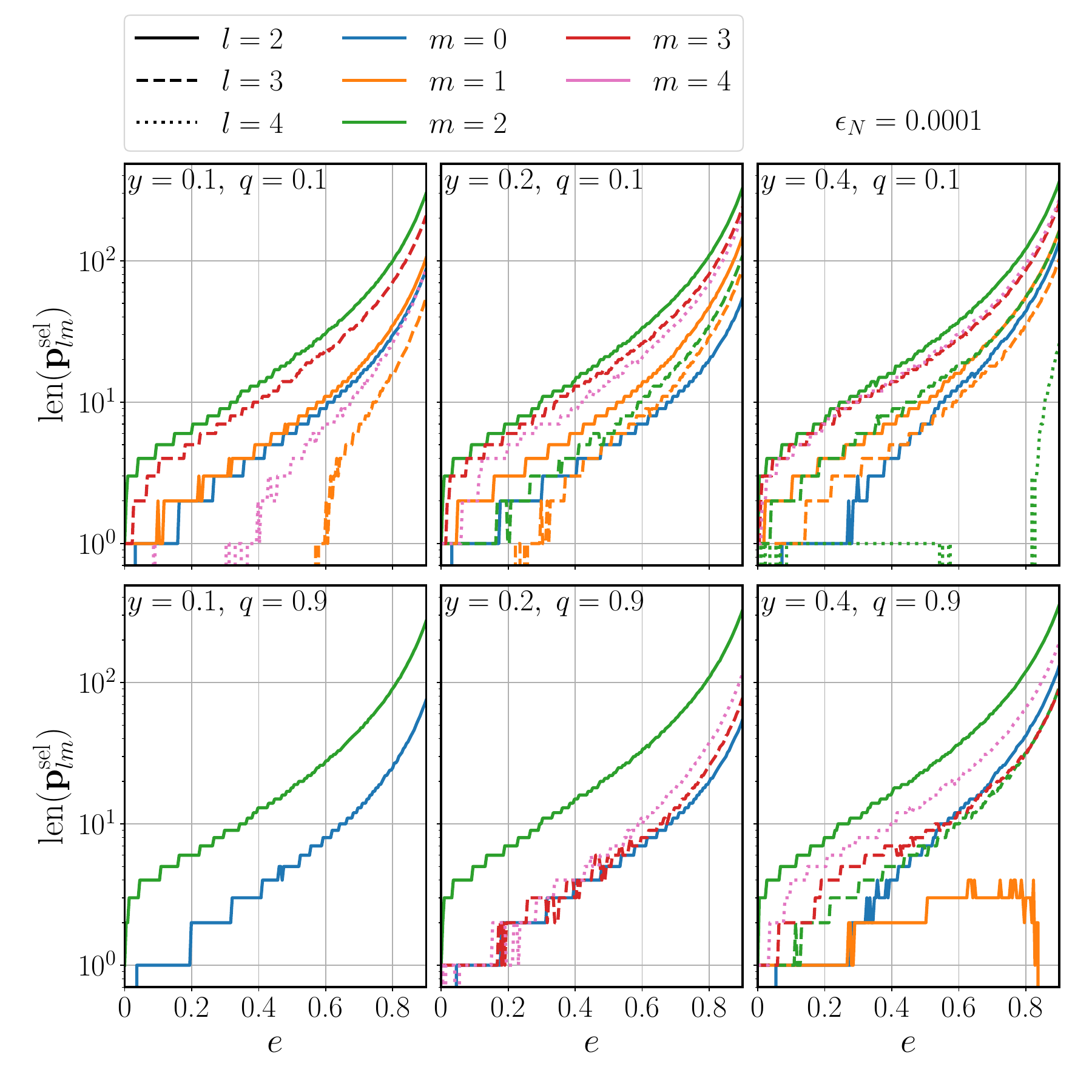}
\caption{\justifying Optimal number of Fourier modes $\mathrm{len}(\bm{p}_{l m}^\mathrm{sel})$ needed to represent each 1PN GW mode as a function of eccentricity $e$. Each panel shows $\mathrm{len}(\bm{p}_{l m}^\mathrm{sel})$ for values of the PN parameter $y$ and mass ratio $q = m_2/m_1$ matching the configuration in Fig.~\ref{fig:norms_Hlm}. We use an amplitude tolerance of $\epsilon_N = 10^{-4}$, typical in data analysis applications~\cite{Morras:2025nlp}. To find $\bm{p}_{l m}^\mathrm{sel}$ as described around Eq.~\eqref{eq:strain_error_ineq}, we compute the Fourier mode amplitudes $N^{l m}_p$ with Eq.~\eqref{eq:Np} and the 1PN norms of the GW modes with Eq.~\eqref{eq:ezNorms}, except for the $(2,2)$ mode, for which we use Eq.~\eqref{eq:normH22}.}
\label{fig:num_Fourier_Modes_Needed}
\end{figure}

In Fig.~\ref{fig:num_Fourier_Modes_Needed} we show the optimal number of Fourier modes that have to be included for each 1PN GW mode, $\mathrm{len}(\bm{p}_{l m}^\mathrm{sel})$, as a function of eccentricity $e$, and for different values of the PN parameter $y$ and mass ratio $q = m_2/m_1$. We use an amplitude tolerance of $\epsilon_N = 10^{-4}$, typical in data analysis applications~\cite{Morras:2025nlp}. Comparing Fig.~\ref{fig:num_Fourier_Modes_Needed} with Fig.~\ref{fig:Nguess_Hlm}, we observe that, as was expected, the number of Fourier modes required to accurately represent the strain is much smaller than Eq.~\eqref{eq:nlm_guess} suggest. Moreover, the relative number of the Fourier modes for each GW mode is also different, with the relative number of $(3,3)$ and $(4,4)$ modes being enhanced, while the $(2,0)$ mode is suppressed compared to earlier estimates (see Fig.~\ref{fig:Nguess_Hlm}).

\begin{figure}[h!]
\centering  
\includegraphics[width=0.5\textwidth]{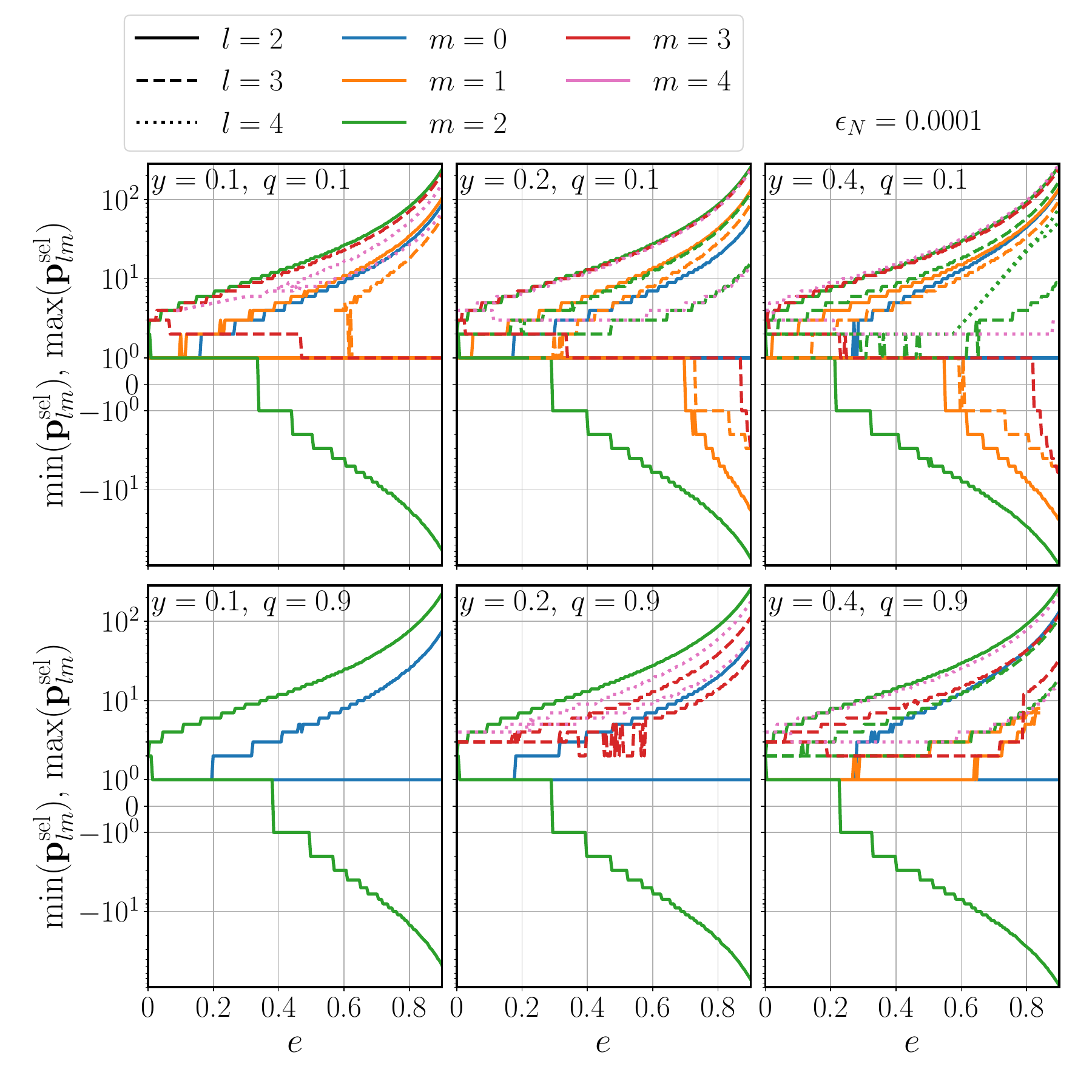}
\caption{\justifying Minimum, $\mathrm{min}(\bm{p}_{l m}^\mathrm{sel})$, and maximum $\mathrm{max}(\bm{p}_{l m}^\mathrm{sel})$ values of $p$ for the optimal Fourier modes that have to be included for each 1PN GW mode, as a function of eccentricity $e$. The selected Fourier modes $\bm{p}_{l m}^\mathrm{sel}$ are the same as in Fig.~\ref{fig:num_Fourier_Modes_Needed}.}
\label{fig:pminmax_Fourier_Modes_Needed}
\end{figure}

To further explore the properties of the optimal included Fourier modes, in Fig.~\ref{fig:pminmax_Fourier_Modes_Needed} we show the largest and smallest $p$ that is included for each 1PN GW mode, with the same tolerance, PN parameters and mass ratios as in Fig.~\ref{fig:num_Fourier_Modes_Needed}. We find that modes with negative $p$ are generally suppressed. The only notable exceptions are the $(2,2)$ mode and, to a lesser extent, the $(2,1)$, $(3,1)$, and $(3,3)$ modes at large $y$ and extreme mass ratios. Interestingly, at high $y$ and extreme $q$, the highest $p$ values for the $(3,3)$ and $(4,4)$ modes approach those of the dominant $(2,2)$ mode, even though their contributions to the strain are much smaller (see Fig.~\ref{fig:norms_Hlm}).

These behaviors can be understood using the asymptotic expansion of $N^{l m}_p$ as $|p| \to \pm \infty$, derived in App.~\ref{sec:appendix:Asymptotic}, where we find

\begin{align}
    N^{l m}_p \xrightarrow[p \to \pm \infty]{}  \kappa^{l m}_\pm  |p|^{n^{l m}_\pm - \frac{1}{2}} \rme^{-\alpha(e) |p|} \left\{1 + \ord{\frac{1}{p}} \right\} ,
    \label{eq:NlmpAsymp}
\end{align}

\noindent with $n^{l m}_\pm$ the constant exponent of the leading power of $p$ in the corresponding formula of Eqs.~(\ref{eq:NpAsympPlus},\ref{eq:NpAsympNeg}), $\kappa^{l m}_\pm $ the prefactor, and we have defined

\begin{equation}
    \alpha(e) = \log\left( \frac{1 + \sqrt{1-e^2}}{e} \right)  - \sqrt{1-e^2} \, .
    \label{eq:DebyeExp}
\end{equation}

Therefore, while the exponential decay in Eq.~\eqref{eq:NlmpAsymp} is the same for all GW modes, the ones with larger $n^{l m}_\pm$ will have $N^{l m}_p$ distributions with heavier tails, which require including more Fourier modes. Since $n^{3 3}_{+} = 2$ and $n^{4 4}_+ = 3$ are among the largest values of $n^{l m}_\pm$, very high frequencies need to be included for the these modes. In Eqs.~(\ref{eq:NpAsympPlus},\ref{eq:NpAsympNeg}) we observe that for $m > 0$, $n^{l m}_{+} \gg n^{l m}_{-}$, explaining why the number of positive frequency Fourier modes included is so much larger than the negative one. An exception to this are the $(2,2)$, $(2,1)$ and $(3,1)$ modes, which have $n^{2 2}_- = n^{2 1}_- = n^{3 1}_- = 0$, explaining why negative frequency Fourier modes are important in these cases.

The asymptotic expansion of $N^{l m}_p$ can also be used to obtain an estimate of the Fourier modes that have to be included. We can expect this approach to work well, since the strain error of Eq.~\eqref{eq:strain_error_def} is mostly due to the Fourier modes neglected in the large $|p|$ tails. This error can be estimated by summing $|N^{l m}_p|^2$ from the maximum selected $|p|$ to $\infty$. Given Eq.~\eqref{eq:NlmpAsymp}, this is related to

\begin{align}
    S_{n,p_0}(\alpha) =& \sum_{p = p_0}^\infty p^{2 n - 1} \rme^{-2 p \alpha} \nonumber \\
    =& p_0^{2 n - 1} \rme^{-2 p_0 \alpha} \sum_{q = 0}^\infty \left(1 + \frac{q}{p_0} \right)^{2 n - 1}  \rme^{-2 q \alpha} \nonumber \\
    =& \frac{p_0^{2 n - 1} \rme^{-2 p_0 \alpha}}{1 - \rme^{-2 \alpha}}\left\{1 + \frac{2 n - 1}{\rme^{2 \alpha} - 1} \frac{1}{p_0} + \ord{\frac{1}{p_0^2}}  \right\} \, .
    \label{eq:Sp0n_alpha}
\end{align}

For such an error term, we can estimate the value of $p$, such that $S_{n,p_n(\alpha,\epsilon)}(\alpha) = \epsilon$ for $\epsilon \ll 1$, as

\begin{align}
    p_n(\alpha,\epsilon) =& p^\mathrm{LO}(\alpha,\epsilon) + \frac{2 n - 1}{2 \alpha}  \log\left\{ 1 +  p^\mathrm{LO}(\alpha,\epsilon) \right\} \nonumber\\
    & + \ord{\frac{\log\left\{p^\mathrm{LO}(\alpha,\epsilon) \right\}}{p^\mathrm{LO}(\alpha,\epsilon)}} \, ,
    \label{eq:p_n_a_eps}
\end{align}    

\noindent where we have introduced $p^\mathrm{LO}(\alpha,\epsilon)$ as the leading order solution, given by

\begin{equation}
    p^\mathrm{LO}(\alpha,\epsilon) = -\frac{\log\left\{(1 - \rme^{-2 \alpha}) \epsilon \right\}}{2 \alpha}
    \label{eq:pLO_n_a_eps}
\end{equation}

\noindent which is a large parameter when $\epsilon \ll 1$, justifying the expansion of Eq.~\eqref{eq:p_n_a_eps}.

\begin{figure}[h!]
\centering  
\includegraphics[width=0.5\textwidth]{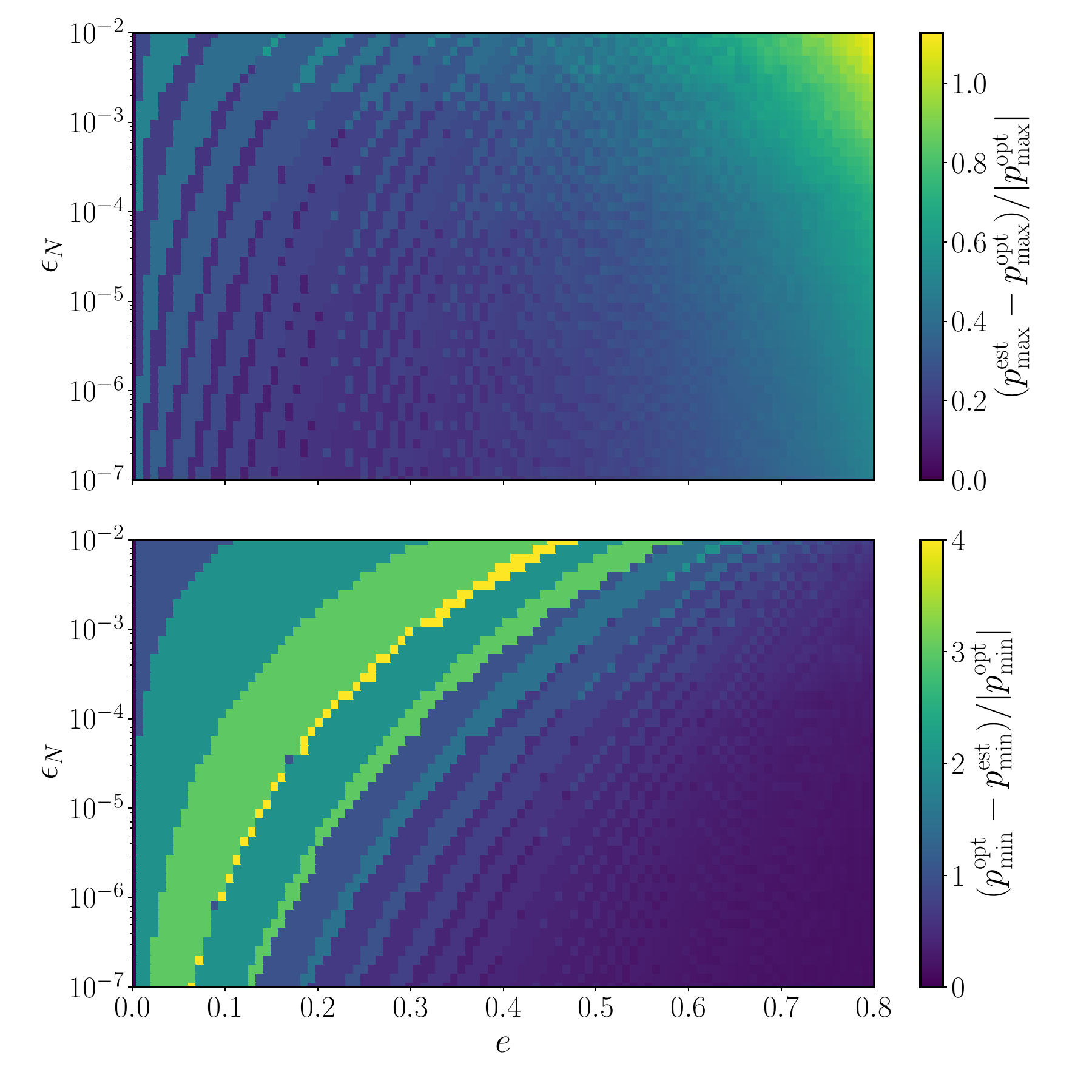}
\caption{\justifying  Relative error between optimal maximum and minimum $p$ values ($p^\mathrm{opt}_\mathrm{max}$, $p^\mathrm{opt}_\mathrm{min}$) and their corresponding estimates ($p^\mathrm{est}_\mathrm{max}$, $p^\mathrm{est}_\mathrm{min}$). The optimal values are obtained as $p^\mathrm{opt}_\mathrm{max} = \max_{l,m,y,q} \bm{p}^\mathrm{sel}_{l m}$ and $p^\mathrm{opt}_\mathrm{min} = \min_{l,m,y,q} \bm{p}^\mathrm{sel}_{l m}$, where we extremize over $y \in [0, 6^{-1/2}]$ and $q = m_2/m_1 \in [0,1]$. The estimated values are computed using Eq.~\eqref{eq:pest_asymp}, which when including all 1PN GW modes, as is the case in this figure, become $p^\mathrm{est}_\mathrm{max} = \lceil 4 + p_{3}(\alpha(e),\epsilon_N) \rceil$ and $p^\mathrm{est}_\mathrm{min} = \lfloor 1 - p_{0}(\alpha(e),\epsilon_N)\rfloor$.}
\label{fig:popt_pest}
\end{figure}

In Fig.~\ref{fig:popt_pest} we show how $p_n(\alpha,\epsilon)$ can be used to make simple yet accurate, estimates for the maximum and minimum values of $p$ that are required when optimally determining the optimal Fourier modes to be included. In particular, we estimate these as

\begin{subequations}
\label{eq:pest_asymp}
\begin{align}
    p_\mathrm{max}^\mathrm{est} =& \Big\lceil |m|_\mathrm{max} + p_{n^\mathrm{max}_{+}}(\alpha,\epsilon_N) \Big\rceil  \, \\ 
    \nonumber & \\
    p_\mathrm{min}^\mathrm{est} =& \Big\lfloor \max(|m|_\mathrm{min},1) - p_{n^\mathrm{max}_{-}}(\alpha,\epsilon_N) \Big\rfloor \, 
\end{align}
\end{subequations}

\noindent where $|m|_\mathrm{max}$ and $|m|_\mathrm{min}$ are the largest and smallest $|m|$ values among the GW modes considered, while $n^\mathrm{max}_{\pm}$ are the maximum values of $n^{l m}_\pm$. In Fig.~\ref{fig:popt_pest}, we compare these estimates with the maximum and minimum values of $p$ that have to be included when using the optimal algorithm described in this section, extremized over $l$, $m$, $q \in [0,1]$ and $y \in [0, 6^{-1/2}]$. We find that $p_\mathrm{max}^\mathrm{est} \geq p_\mathrm{max}^\mathrm{opt}$ and $p_\mathrm{min}^\mathrm{est} \leq p_\mathrm{min}^\mathrm{opt}$ in all cases, confirming that Eq.~\eqref{eq:pest_asymp} provides simple upper and lower bounds for the values of $p$ that have to be explored in order to optimize the Fourier modes included. Since 

\begin{equation}
    \alpha(e) \xrightarrow[e \to 1]{} \frac{1}{3}(1-e^2)^{3/2} \, ,
    \label{eq:alpha_lim}
\end{equation}

\noindent both $p_\mathrm{max}^\mathrm{est}$ and $p_\mathrm{min}^\mathrm{est}$ scale as $(1-e^2)^{-3/2}$ when $e \to 1$. This reinforces the result in Sec.~\ref{sec:Properties:Estimation}, where we found an identical scaling for the simple estimation of the number of required Fourier modes.

Finally, in Fig.~\ref{fig:popt_pest} we also observe that the estimates become increasingly accurate as $\epsilon_N \to 0$, consistent with the asymptotic nature of the approximation. Importantly, both $p_\mathrm{max}^\mathrm{est}$ and $p_\mathrm{min}^\mathrm{est}$ scale as $\log(1/\epsilon_N)$ when $\epsilon_N \to 0$, meaning that highly accurate strain representations can be achieved with only a modest increase in the number of Fourier modes.

\section{Future extensions}
\label{sec:future}

The choice to compute the amplitudes at 1PN order in this paper was deliberate, as expressions become significantly more complex at 1.5PN order and beyond. Furthermore, these higher-order PN corrections are expected to have a small effect on the strain and are unlikely to be observable with current detectors~\cite{Mills:2020thr}. The main drawback of truncating at 1PN is that the amplitudes lose accuracy earlier in the inspiral, as higher-PN terms become increasingly important closer to merger. Nonetheless, in this section, we discuss the additional ingredients required to extend the PN order of the Fourier mode amplitudes, outlining how some parts can be computed and the challenges that are expected to arise. We note that including tail terms, spin contributions, and higher PN corrections is likely to improve the inspiral description, delaying the breakdown of the approximation and providing a more faithful comparison to numerical relativity~\cite{Boyle:2007ft,Hannam:2007wf,Hannam:2007ik}, although we defer a quantitative study of these effects to future work.

\subsection{Tail effects}
\label{sec:future:tails}

These are nonlinear, hereditary effects that arise due to the backscattering of GWs off the spacetime curvature generated by the source itself~\cite{Blanchet:1992br}. These effects are a key prediction of General Relativity, and, even if their contribution starts at 1.5PN order, they can be numerically significant compared to instantaneous (non-tail) terms. The leading-order tail contribution is~\cite{Blanchet:1992br}

\begin{align}
    H^{l m}_\mathrm{tail} = & 2 M \int_0^\infty \d \tau \left[\log\left(\frac{\tau}{2 \tau_0}\right) + c_{l m} \right] \ddot{H}^{l m}_\mathrm{inst}(t - \tau) \nonumber \\
    & \times \left\{1 + \ord{y^2} \right\} \, ,
    \label{eq:TailIntegral_hlm}
\end{align}

\noindent where $H^{l m}_\mathrm{inst}$ is the instantaneous part of the strain, $c_{l m}$ is a numerical constant (e.g., $c_{2 2} = c_{2 0} = 11/12$), and $\tau_0$ is an arbitrary constant with units of time.

Substituting the Fourier expansion of the strain from Eq.~\eqref{eq:Hlm_Nlmp} into Eq.~\eqref{eq:TailIntegral_hlm}, and performing the change of variables $\tau \to \ell/n$, we obtain

\begin{align}
    (N^{l m}_p)_\mathrm{tail} = & - 2 p^2 (1 - e^2)^{3/2} y^3 (N^{l m}_p)_\mathrm{inst} \left\{1 + \ord{y^2} \right\} \nonumber \\ & \times \int_0^\infty \d \ell \left[\log\left( \frac{\ell}{2 n \tau_0} \right) + c_{l m} \right] \rme^{\rmi p \ell} \, ,
    \label{eq:TailIntegral_Nlmp}
\end{align}

\noindent where $n$ is given to 1PN order in Eq.~\eqref{eq:qKvars:n}. The integrals in Eq.~\eqref{eq:TailIntegral_Nlmp} are well known in the literature~\cite{Boetzel:2019nfw}, and evaluating them yields

\begin{align}
    & (N^{l m}_p)_\mathrm{tail} = p (1-e^2)^{3/2} y^3  (N^{l m}_p)_\mathrm{inst} \bigg\{\pi \, \mathrm{sign}(p)  \nonumber\\ 
    & + \rmi \left[ 3  \log{\frac{(1-e^2) y^2}{x_0}} + 2 \log{\frac{|p|}{2}} + \frac{11}{6} - 2 c_{l m} \right] + \ord{y^2} \bigg\}   ,
    \label{eq:Nlmptail_Nlmpinst} 
\end{align}

\noindent where

\begin{equation}
    x_0 = \left( \frac{M \rme^{11/12 - \gamma_E}}{4 \tau_0} \right)^{2/3} \, ,
    \label{eq:x0_def}
\end{equation}

\noindent and $\gamma_E = 0.577\ldots$ is the Euler-Mascheroni constant. Therefore, using the closed-form analytical expressions for the leading-order $(N^{l m}_p)_\mathrm{inst}$ developed in this paper, we automatically obtain closed-form expressions for the leading-order tail contributions. These results are valid for arbitrary eccentricity, in contrast to the low-eccentricity expansions commonly used in the literature~\cite{Boetzel:2019nfw,Khalil:2021txt,Henry:2023tka}, which can be recovered by Taylor expanding Eq.~\eqref{eq:Nlmptail_Nlmpinst} around $e=0$. 

This Fourier expansion method can be extended to compute higher-order hereditary contributions, yielding simple expressions for the Fourier amplitudes, including effects such as tails-of-tails. This provides a systematic approach to incorporating these corrections without relying on low-eccentricity expansions.

\subsection{Spin effects}
\label{sec:future:spin}

The presence of spin in the binary components leaves an imprint on the amplitude of the observed strain. This effect is important to include, as it can help break parameter degeneracies. Moreover, for large component spins, this contributions can become numerically significant compared to non-spinning terms. The leading-order spin corrections to the waveform amplitudes arise from spin-orbit interactions and scale as

\begin{align}
    (K^{l m})_\mathrm{spin} \sim 
    \begin{cases}
        \ord{y^{l+1}} &, \; \text{if $l+m$ is even} \\
        \ord{y^{l}} &, \; \text{if $l+m$ is odd}  \\
     \end{cases}
    \, .
    \label{eq:Klm_spin_PNorder}
\end{align}

Comparing with Eq.~\eqref{eq:Klm_PN_order}, the relative PN order of spin corrections is 1.5PN for modes with $l + m$ even, and 0.5PN for modes with $l + m$ odd. At 1.5PN order, spin terms must be included for the $(2,0)$, $(2,2)$, $(3,0)$, and $(3,2)$ modes, and these corrections are listed in App.~\ref{sec:appendix:Spin}. For modes with $l + m$ odd, closed-form expressions for the leading-order spin terms in the Fourier mode amplitudes can be obtained, since corrections to the Keplerian orbits can be neglected at this order. In contrast, for modes with $l + m$ even, spin effects require incorporating 1.5PN corrections in the quasi-Keplerian parametrization. As a result, the same class of rapidly converging Bessel series that appears in the 1PN instantaneous terms may also arise here. This can be explicitly seen in Eq.~\eqref{eq:NpSpin22} for the $(2,2)$ mode.

\subsection{Higher-order instantaneous and quasi-Keplerian corrections}
\label{sec:future:HigherPN}

At higher PN orders, we must not only include corrections to the $(l,m)$ modes studied in this paper, but also account for additional modes. From Eq.~\eqref{eq:Klm_PN_order}, at 1.5PN we must include the $(4,1)$, $(4,3)$, $(5,1)$, $(5,3)$, and $(5,5)$ modes. While such modes can, in principle, contribute additional information to the waveform, potentially helping to break parameter degeneracies, in practice they have such small amplitudes that they cannot be observed by current detectors and are typically ignored in waveform models~\cite{Garcia-Quiros:2020qpx,Thompson:2023ase,Albanesi:2025txj}. An exception to this are the $(4,3)$ and $(5,5)$ modes, which are sometimes included~\cite{Gamboa:2024hli,Planas:2025feq}, as they have the largest contributions among the 1.5PN modes. As with the modes computed in this work, the leading-order Fourier amplitudes of any additional modes can be derived analytically, since they depend only on the positions and velocities of a Keplerian binary~\cite{Blanchet:1998in}, which admit closed-form Fourier expansions.

However, at higher PN orders we must also incorporate corrections to the quasi-Keplerian parametrization, which enter at 1PN relative order. At 1.5PN, these corrections must be included for the $(2,1)$, $(3,1)$, and $(3,3)$ modes. As in the case of the $(2,2)$ mode studied in this paper, a substantial portion of these corrections can be integrated analytically, while the remaining terms can be expressed as rapidly converging series involving Bessel functions, similar to those in Eq.~\eqref{eq:fbetafourier}. Extending this approach to include corrections at 2PN relative order is expected to involve significantly more complex integrals, and we leave their detailed study to future work.

\section{Conclusions}
\label{sec:conclusions}

In this work, we have studied the GW amplitudes emitted by inspiraling eccentric binaries. In particular, we derived simple expressions for the 1PN Fourier amplitudes of the $(l,m)$ modes contributing at this order, valid for arbitrary eccentricities. We also developed tools to characterize the GW emission of eccentric binaries, computing the contribution of each $(l,m)$ mode to the total strain, its mean frequency, frequency spread, and asymptotic behavior at large frequencies. Additionally, we developed a method to optimally truncate the Fourier series expansion at a given accuracy, minimizing the computational cost of waveform generation. Finally, we discussed how our method can be extended to higher PN orders, showing that it can be used to obtain closed-form expressions for the leading order tail and spin effects, and outlining the steps required to include higher-order corrections.

The results presented in this paper can improve both the accuracy and efficiency of eccentric waveform models, particularly for systems with high orbital eccentricity. The accurate modeling of subleading $(l,m)$ modes presented in this work is especially important for breaking parameter degeneracies and avoiding systematic biases in parameter estimation~\cite{VanDenBroeck:2006ar,Varma:2014jxa,Mills:2020thr}. Moreover, because our formalism is based on a spherical harmonic decomposition, it can be readily extended to include spin-precession effects using the “twisting-up” approximation~\cite{Schmidt:2010it,OShaughnessy:2011pmr,Boyle:2011gg,Schmidt:2012rh,Klein:2021jtd}. Finally, we showed how the techniques introduced in this paper can be extended to higher PN orders, which may be required to model the high signal-to-noise ratio events expected in future GW observatories.

\section*{Code Availability}

A repository containing python scripts and Mathematica notebooks to reproduce the formulas and figures in this paper is available at Ref.~\cite{EccentricHMs_repo}. 

\section*{Acknowledgments}

I thank Geraint Pratten for helpful feedback as internal reviewer for LIGO and Virgo.
G.M. acknowledges support from the Ministerio de Universidades through Grant No. FPU20/02857, from the Agencia Estatal de Investigaci\'on through the Grant IFT Centro de Excelencia Severo Ochoa No. CEX2020-001007-S, funded by MCIN/AEI/10.13039/501100011033, and from grant PID2021-123012NB-C43 [MICINN-FEDER]
This manuscript has the LIGO document number P2500396.


\FloatBarrier
\onecolumngrid
\appendix

\section{Expressions for $K^{l m}$}
\label{sec:appendix:Klm}

In this appendix, we provide the 1PN expressions for $K^{l m}$, extracted from Refs.~\cite{Mishra:2015bqa,Khalil:2021txt} and converted to our notation. That is

\begin{subequations}
\label{eq:K}
\begin{align}
K^{2 0} =& \sqrt{\frac{2}{3}} e \Bigg\{ \frac{\cos{u}}{1-e \cos{u}} + \frac{y^2}{7 (1-e \cos{u})^3} \Bigg[ e \left(1-e^2\right) (26-\nu ) + \left(-\frac{61}{2}+\frac{33}{2} e^2 -\frac{11}{6} \left(1-e^2\right) \nu \right) \cos{u} \nonumber \\
& + e \left(9+19 e^2 + \frac{17}{3} \left(1-e^2\right) \nu \right) \cos^2{u}
   \left(1-\frac{e}{2} \cos{u}\right)\Bigg] \Bigg\} \label{eq:K:20} \, , \\
K^{2 1} =& \frac{2}{3} \rmi y \left(\delta \mu - \frac{3}{2} y \delta\chi \right)  \frac{1-e^2}{(1-e \cos{u})^2} \label{eq:K:21} \, , \\
K^{2 2} =& \frac{2 \left(1-e^2\right)-e\cos{u} + e^2 \cos^2{u} + 2 \rmi e \sqrt{1-e^2} \sin{u}}{(1-e \cos{u})^2} - \frac{y^2}{7 (1-e \cos{u})^3} \Bigg\{ \frac{1}{3} \left(1-e^2\right) \left(107-55 \nu +e^2 (139-32 \nu )\right) \nonumber \\
& + \frac{e}{2} \left[-135+41 \nu +e^2 \left(69-\frac{89 \nu }{3}\right)+e^4 \left(38-\frac{34 \nu }{3}\right)\right] \cos{u} + e^2 \left[9+19 e^2+\frac{17}{3} \left(1-e^2\right) \nu \right] \cos^2{u} \left(1-\frac{e}{2} \cos{u}\right) \nonumber \\
& -\rmi e \sqrt{1-e^2} \sin{u} \left[\frac{1}{3} \left(10+19 \nu +e^2 (-136+23
   \nu )\right)+e \left(23-\frac{25 \nu }{3}+e^2 \left(19-\frac{17 \nu
   }{3}\right)\right) \cos{u}\right] \Bigg\} \label{eq:K:22} \, , \\
K^{3 0} =& \label{eq:K:30} -\frac{y^2 (1-3 \nu ) \left(1-e^2\right)^{3/2}}{\sqrt{42} (1-e \cos{u})^3} \rmi e \sin{u} \, , \\
K^{3 1} =& \frac{y \delta \mu  \sqrt{1-e^2}}{\sqrt{14} (1-e \cos{u})^2} \left\{ \rmi \sqrt{1-e^2} \left(\frac{1}{6}-e \cos{u}\right)-e \sin{u}+\frac{e^2}{2} \sin(2 u) \right\} \label{eq:K:31} \, , \\
K^{3 2} =& \sqrt{\frac{5}{7}} \frac{y^2 (1-3 \nu )
   \left(1-e^2\right)^{3/2}}{6 (1-e \cos{u})^3} \left\{ 4 \sqrt{1-e^2} + \rmi e \sin{u} \right\} \label{eq:K:32} \, , \\
K^{3 3} =& -\sqrt{\frac{5}{42}} \frac{y \delta \mu  \sqrt{1-e^2}}{(1-e \cos{u})^3} \Bigg\{\frac{\rmi}{2} \sqrt{1-e^2} \left[9-5 e^2-7 e \cos{u}+3 e^2 \cos(2u)\right] -e \left(5-\frac{15 e^2}{4}\right) \sin{u} \nonumber \\
& + e^2 \sin (2 u) -\frac{e^3}{4} \sin (3 u) \Bigg\} \label{eq:K:33} \, , \\
K^{4 0} =& \frac{1}{7 \sqrt{2}} \frac{y^2 (1-3 \nu ) \left(1-e^2\right)}{(1-e \cos{u})^3} \left\{ -\frac{e^2}{6}+\left(\frac{1}{6}+\frac{3 e^2}{4}\right) e \cos{u}-e^2 \cos
   (2 u)+\frac{e^3}{4} \cos (3 u) \right\} \label{eq:K:40} \, , \\
K^{4 2} =&  \frac{\sqrt{5}}{21} \frac{y^2 (1-3 \nu ) \left(1-e^2\right)}{(1-e \cos{u})^3} \Bigg\{ \frac{2}{3}-\frac{e^2}{2}-e \left(\frac{13}{6}-\frac{5 e^2}{4}\right) \cos{u} + e^2 \cos (2 u)-\frac{e^3}{4} \cos (3 u) \nonumber \\
& + \rmi e \sqrt{1-e^2} \left(\frac{5}{2}\sin{u} - e \sin (2 u)\right) \Bigg\} \label{eq:K:42} \, , \\
K^{4 4} =& \sqrt{\frac{5}{7}} \frac{y^2 (1-3 \nu ) \left(1-e^2\right)}{(1-e \cos{u})^4} \Bigg\{ -\frac{16}{9}+\frac{173 e^2}{72}-\frac{35 e^4}{48} + e \left(\frac{65}{36}-\frac{37 e^2}{24}\right) \cos{u} + e^2 \left(-\frac{61}{72}+\frac{7 e^2}{12}\right) \cos (2 u) \nonumber \\
& + \frac{e^3}{8} \cos (3 u) -\frac{e^4}{48} \cos (4 u) - \rmi e \sqrt{1-e^2} \left[ \left(\frac{13}{6}-\frac{7 e^2}{6}\right) \sin{u}-\frac{3}{4} e \sin (2 u)+\frac{e^2}{6} \sin (3 u) \right]
\Bigg\}  \label{eq:K:44} \, .
\end{align}
\end{subequations}

\noindent where $\delta \mu = (m_1 - m_2)/(m_1 + m_2) = \sqrt{1 - 4\nu}$ is the asymmetric mass ratio.

\section{Derivative Expressions for $F^{l m}$}
\label{sec:appendix:intFLm}

In this appendix, we provide the 1PN expressions for $F^{l m}$ (defined in Eq.~\eqref{eq:Flm_def}) in terms of derivatives with respect to $\ell$, such that their Fourier transforms are simple to compute with Bessel functions, as explained in Sec.~\ref{sec:1PN}. That is

\begin{subequations}
\label{eq:F}
\begin{align}
F^{2 0} =& \sqrt{\frac{2}{3}} e \Bigg\{ \left[ 1-y^2 \left(\frac{9}{14}+\frac{17 \nu }{42}+e^2 \left(\frac{19}{14}-\frac{17 \nu }{42}\right)\right) \right] \frac{\cos{u}}{1-e \cos{u}}  + \rmi \frac{\d}{\d\ell} \left[ y^2 \left(1-e^2\right) \left(\frac{26}{7}-\frac{\nu }{7}\right) \frac{\rmi \sin{u}}{1-e \cos{u}}  \right] \Bigg\} \label{eq:F:20} , \\
F^{2 1} =& -\frac{2}{3} y \left(\delta \mu - \frac{3}{2} y \delta\chi \right)  \left(1-e^2\right)\frac{\d}{\d\ell} \left[ \frac{\sqrt{1-e^2} \cos{u} - \rmi \sin{u}}{1-e \cos{u}} \right] \label{eq:F:21} \, , \\
F^{2 2} =& \frac{e}{1-e \cos{u}} \Bigg\{ \cos{u}+y^2 \left[\left(-\frac{111}{14}+\frac{39 \nu }{14}+e^2 \left(-\frac{19}{14}+\frac{17 \nu }{42}\right)\right) \cos{u}+ \rmi \sqrt{1-e^2} \left(\frac{37}{7}-\frac{25 \nu }{21}\right) \sin{u}\right] \Bigg\} \nonumber \\
& + \rmi \frac{\d}{\d \ell} \Bigg\{\frac{2 \rmi e \sin{u}-\rmi \sin (2 u)-\sqrt{1-e^2} (1-\cos (2 u))}{1-e \cos{u}} + \frac{y^2}{1-e \cos{u}} \Bigg[ \sqrt{1-e^2} \left(-\frac{115}{14}-\frac{19 \nu }{14}+e^2 \left(\frac{356}{21}-\frac{11 \nu }{21}\right)\right) \nonumber \\
& \qquad\quad + \rmi e \left(-\frac{262}{21}+\frac{65 \nu }{21}+e^2 \left(\frac{23}{21}+\frac{8 \nu }{21}\right)\right) \sin{u}+\left(\frac{37}{14}-\frac{67 \nu}{42}\right) \left(\rmi \sin (2 u)-\sqrt{1-e^2} \cos (2 u)\right) + f_\beta(u,e) \Bigg] \Bigg\} \nonumber \\
& + \frac{\d^2}{\d \ell^2} \Bigg\{ \frac{1-3 \nu }{1-e \cos{u}} \left[ \left(\frac{5}{21}-\frac{2 e^2}{7}+\frac{2 e^4}{21}\right)-\frac{2}{21} \rmi e
   \sqrt{1-e^2} \left(2-e^2\right) \sin{u}-\frac{1}{21} \left(\cos (2 u)-\rmi
   \sqrt{1-e^2} \sin (2 u)\right) \right] \Bigg\}  \label{eq:F:220} \, , \\
F^{3 0} =& \frac{\rmi}{\sqrt{42}} y^2 (1-3 \nu ) \left(1-e^2\right)^{3/2} \frac{\d}{\d\ell} \left[ \frac{1}{1-e \cos{u}} \right] \label{eq:F:30} \, , \\
F^{3 1} =& \frac{\rmi}{\sqrt{14}} y \delta \mu  \sqrt{1-e^2} \Bigg\{ \frac{\sqrt{1-e^2} \cos{u} - \rmi \sin{u}}{1-e \cos{u}} - \rmi \frac{5}{6} \sqrt{1-e^2} \frac{\d}{\d \ell} \left[ \frac{\sqrt{1-e^2} \cos{u} - \rmi \sin{u}}{1-e \cos{u}} \right]  \Bigg\} \label{eq:F:31} \, , \\
F^{3 2} =& \frac{\rmi}{6} \sqrt{\frac{5}{7}} y^2 (1-3 \nu ) \left(1-e^2\right)^{3/2} \Bigg\{\frac{1}{1-e \cos{u}} - \rmi \frac{\d}{\d\ell} \left[ \frac{\sqrt{1-e^2} (1-\cos (2 u))-\rmi (2 e \sin{u}-\sin (2 u))}{1-e \cos{u}} \right] \Bigg\} \label{eq:F:32} \, , \\
F^{3 3} =& \rmi \sqrt{\frac{5}{42}} y \delta \mu  \sqrt{1-e^2} \Bigg\{ \frac{\sqrt{1-e^2} \cos{u}-3 \rmi \sin{u}}{1-e \cos{u}} + \frac{\rmi}{2} \frac{\d}{\d\ell} \left[ \frac{5 e-13 \cos{u}+7 \rmi \sqrt{1-e^2} \sin{u}}{1-e \cos{u}} \right] \nonumber\\
& + \frac{\d^2}{\d\ell^2} \left[ \frac{\sqrt{1-e^2} \left(2 e-\frac{5}{2} \cos{u} +\frac{1}{2} \cos (3 u)\right)+\rmi \left(\frac{7}{2}-2 e^2\right) \sin{u}-\frac{\rmi }{2} \sin (3 u)}{1-e \cos{u}} \right] \Bigg\} \label{eq:F:33} \, , \\
F^{4 0} =& \frac{1}{7 \sqrt{2}} y^2 (1-3 \nu ) \left(1-e^2\right) \Bigg\{ \frac{e \cos{u}}{1-e \cos{u}} + \rmi \frac{5}{6} \frac{\d}{\d\ell} \left[ \frac{\rmi e \sin{u}}{1-e \cos{u}}\right]\Bigg\}\label{eq:F:40} \, , \\
F^{4 2} =& \frac{\sqrt{5}}{21} y^2 (1-3 \nu ) \left(1-e^2\right) \Bigg\{ \frac{e \cos{u}}{1-e \cos{u}} - \rmi \frac{\d}{\d\ell} \left[ \frac{\sqrt{1-e^2} \left(\frac{1}{2}-\cos (2 u)\right)-\frac{7}{6} \rmi e \sin{u} + \rmi \sin (2 u)}{1-e \cos{u}} \right] \nonumber \\
& - \frac{1}{3} \sqrt{1-e^2}\frac{\d^2}{\d\ell^2} \left[ \frac{\sqrt{1-e^2} (1-\cos (2 u))-\rmi (2 e \sin{u}-\sin (2 u))}{1-e \cos{u}} \right]  \Bigg\} \label{eq:F:42} \, , \\
F^{4 4} =& \frac{1}{6} \sqrt{\frac{5}{7}} y^2 (1 - 3 \nu ) \left(1-e^2\right) \Bigg\{ \frac{e \cos{u}}{1-e \cos{u}} - \rmi \frac{\d}{\d\ell} \left[ \frac{\sqrt{1-e^2} (1-2 \cos (2 u))-\frac{5}{6} \rmi e \sin{u}}{1-e \cos{u}} \right] \nonumber \\
& -\frac{1}{3} \sqrt{1-e^2} \frac{\d^2}{\d\ell^2} \left[ \frac{2 \sqrt{1-e^2} (1-3 \cos (2 u))-\rmi (10 e \sin{u}-11 \sin (2 u))}{1-e \cos{u}} \right] \nonumber \\
& + \frac{\rmi}{6} \sqrt{1-e^2} \frac{\d^3}{\d\ell^3} \left[ \frac{15-8 e^2-8 \cos (2 u)+\cos (4 u)-\rmi \sqrt{1-e^2} (8 e \sin{u}-6 \sin (2 u)+\sin (4 u))}{1-e \cos{u}} \right] \Bigg\} \label{eq:F:44} \, ,
\end{align}
\end{subequations}

\noindent where in $F^{22}$ we have defined

\begin{subequations}
\label{eq:fbeta}
\begin{align}
f_\beta(u,e) =& \frac{18 e^4+36 \rmi e \sin{u} - 24 e^3 \rme^{\rmi u}}{1+\sqrt{1-e^2}} + \frac{6 e \rme^{-\rmi u}}{\left(1+\sqrt{1-e^2}\right)^2} \nonumber\\
& - \left[ \frac{9}{2} \frac{e^2}{ 1+\sqrt{1-e^2}}+\frac{6 e^3 \rme^{\rmi u}}{\left(1+\sqrt{1-e^2}\right)^2}+3 \left(\left(1-\frac{e^2}{2}\right) \cos (2 u) - \rmi \sqrt{1-e^2} \sin (2 u)\right) \right] \left(1-e \cos{u}\right) \nonumber\\
& + \frac{12 \left(1-e^2\right)^2}{\left(1+\sqrt{1-e^2}\right)^2} \frac{\rme^{-2 \rmi u}}{1-\frac{e}{1+\sqrt{1-e^2}} \rme^{-\rmi u}} \nonumber \\
& +6 \rmi \left[ \sqrt{1-e^2} (e \cos{u}-\cos (2 u))-\rmi \left(e \sin{u}-\left(1-\frac{e^2}{2}\right) \sin (2 u)\right) \right] \left[2 \arctan\left(\sqrt{\frac{1+e}{1-e}} \tan \frac{u}{2}\right)-u \right] \label{eq:fbeta:exact} \\
=&  \frac{3 \beta }{\left(1+\beta ^2\right)^2} \Bigg\{ \frac{\beta ^3 \left(53+4 \beta ^2-\beta ^4\right)}{1+\beta ^2}+\frac{6-\frac{41 \beta ^2}{3}-\frac{56 \beta ^4}{3}}{1+\beta ^2} \rme^{\rmi u} -\frac{6+13 \beta ^2+7 \beta ^4+\frac{2 \beta ^6}{3}-\frac{\beta ^8}{3}}{1+\beta ^2} \rme^{-\rmi u} \nonumber\\
& -\frac{\beta^3}{6} \rme^{2 \rmi u}-\beta  \left(8-6 \beta ^2+\frac{8 \beta ^4}{3}-\frac{\beta ^6}{2}\right) \rme^{-2 \rmi u}+\frac{\beta ^4}{1+\beta ^2}\rme^{3 \rmi u}+\frac{1}{1+\beta ^2}\rme^{-3 \rmi u} \nonumber\\
&+ \sum_{n=3}^\infty \beta ^{n-3} \left[\beta^4\frac{ 24}{4-5 n^2+n^4} \rme^{\rmi n u} +\left((1 - \beta^2)^4+\frac{2}{n-2}-\frac{4 \beta^2}{n-1}+\frac{4 \beta ^6}{n+1}-\frac{2 \beta ^8}{n+2}\right) \rme^{-\rmi n u}\right] \Bigg\} \, , \label{eq:fbeta:sum}
\end{align}
\end{subequations}

\noindent with

\begin{equation}
    \beta = \frac{e}{1 + \sqrt{1 - e^2}} \, ,
    \label{eq:beta_def}
\end{equation}

\noindent and to go from Eq.~\eqref{eq:fbeta:exact} to Eq.~\eqref{eq:fbeta:sum} we have used that~\cite{Konigsdorffer:2006zt}

\begin{equation}
    2 \arctan\left(\sqrt{\frac{1+e}{1-e}} \tan \frac{u}{2}\right) - u = 2 \arctan\left(\frac{\beta \sin{u}}{1 - \beta \cos{u}}\right) = \rmi \log\left( \frac{1 - \beta \rme^{\rmi u}}{1 - \beta \rme^{-\rmi u}} \right) = 2 \sum_{n=1}^\infty \frac{\beta^n}{n} \sin{(n u)} \, .
    \label{eq:arctan_beta}
\end{equation}

\section{Expressions for Fourier Mode Amplitudes}
\label{sec:appendix:FourierModes}

In this appendix, we provide the 1PN expressions of $N^{l m}_p$, computed with Eq.~\eqref{eq:Nlm_p} as the Fourier series coefficients of the $F^{l m}$ listed in Eq.~\eqref{eq:F}. For simplicity, we write them in terms of

\begin{subequations}
\label{eq:Jsim}
\begin{align}
\mathcal{C}_{n,p}(z) =& J_{p+n} (z) + J_{p-n} (z) \, , \label{eq:Jsim:C} \\
\mathcal{S}_{n,p}(z) =& J_{p+n} (z) - J_{p-n} (z) \, , \label{eq:Jsim:S}
\end{align}
\end{subequations}

\noindent to obtain

\begin{subequations}
\label{eq:Np}
\begin{align}
N^{2 0}_p =& \sqrt{\frac{1}{6}} e \left\{\left[1-y^2 \left(\frac{9}{14}+\frac{17 \nu}{42}+e^2 \left(\frac{19}{14}-\frac{17 \nu }{42}\right)\right)\right] \mathcal{C}_{1,p}(p e) + p y^2 \left(1-e^2\right) \left(\frac{26}{7}-\frac{\nu}{7}\right) \mathcal{S}_{1,p} (p e) \right\} \label{eq:Np:20} \, , \\
N^{2 1}_p =& \frac{\rmi}{3} p  y \left(\delta \mu - \frac{3}{2} y \delta\chi \right)   \left(1-e^2\right) \left(\sqrt{1-e^2} \mathcal{C}_{1,p}(p e)-\mathcal{S}_{1,p}(p e)\right) \label{eq:Np:21} \, , \\
N^{2 2}_p =& \frac{e}{2} \left\{\mathcal{C}_{1,p}(p e)+y^2 \left[\left(-\frac{111}{14}+\frac{39 \nu }{14}+e^2 \left(-\frac{19}{14}+\frac{17 \nu }{42}\right)\right) \mathcal{C}_{1,p}(p e)+\sqrt{1-e^2} \left(\frac{37}{7}-\frac{25 \nu}{21}\right) \mathcal{S}_{1,p}(p e)\right]\right\} \nonumber \\
& + p \Bigg\{e \mathcal{S}_{1,p}(p e)-\frac{1}{2} \mathcal{S}_{2,p}(p e)-\frac{1}{2} \sqrt{1-e^2} (\mathcal{C}_{0,p}(p e)-\mathcal{C}_{2,p}(p e)) \nonumber \\
& + y^2 \Bigg[\sqrt{1-e^2} \left(-\frac{115}{28}-\frac{19 \nu }{28}+e^2 \left(\frac{178}{21}-\frac{11 \nu }{42}\right)\right) \mathcal{C}_{0,p}(p e)+e \left(-\frac{131}{21}+\frac{65 \nu }{42}+e^2 \left(\frac{23}{42}+\frac{4 \nu }{21}\right)\right) \mathcal{S}_{1,p}(p e) \nonumber \\
& +\left(\frac{37}{28}-\frac{67 \nu}{84}\right) \left(\mathcal{S}_{2,p}(p e)-\sqrt{1-e^2} \mathcal{C}_{2,p}(p e)\right) + \tilde{f}_{\beta,p}(e) \Bigg]\Bigg\} - y^2 p^2 (1-3 \nu ) \Bigg\{\left(\frac{5}{42} - \frac{e^2}{7} +\frac{e^4}{21}\right) \mathcal{C}_{0,p}(p e) \nonumber \\
& -\frac{1}{21} e \sqrt{1-e^2} \left(2-e^2\right) \mathcal{S}_{1,p}(p e)-\frac{1}{42} \left(\mathcal{C}_{2,p}(p e)-\sqrt{1-e^2} \mathcal{S}_{2,p}(p e)\right)\Bigg\} \label{eq:Np:22} \, , \\
N^{3 0}_p =& \frac{1}{2 \sqrt{42}} p y^2 (1-3 \nu ) \left(1-e^2\right)^{3/2} \mathcal{C}_{0,p}(p e) \label{eq:Np:30} \, , \\
N^{3 1}_p =& \frac{\rmi}{2 \sqrt{14}} y \delta \mu  \sqrt{1-e^2} \left(1-\frac{5}{6} p \sqrt{1-e^2}\right) \left(\sqrt{1-e^2} \mathcal{C}_{1,p}(p e)-\mathcal{S}_{1,p}(p e)\right)\label{eq:Np:31} \, , \\
N^{3 2}_p =& \frac{1}{12} \sqrt{\frac{5}{7}} p y^2 (1-3 \nu ) \left(1-e^2\right)^{3/2} \left\{\mathcal{C}_{0,p}(p e)-p \left[\sqrt{1-e^2} (\mathcal{C}_{0,p}(p e)-\mathcal{C}_{2,p}(p e))-(2 e \mathcal{S}_{1,p}(p e)-\mathcal{S}_{2,p}(p e))\right]\right\} \label{eq:Np:32} \, , \\
N^{3 3}_p =& \frac{\rmi}{2} \sqrt{\frac{5}{42}} y \delta \mu  \sqrt{1-e^2} \Bigg\{\sqrt{1-e^2} \mathcal{C}_{1,p}(p e)-3 \mathcal{S}_{1,p}(p e)+\frac{p}{2} \left[5 e \mathcal{C}_{0,p}(p e)-13 \mathcal{C}_{1,p}(p e)+7 \sqrt{1-e^2} \mathcal{S}_{1,p}(p e)\right] \nonumber\\
& - p^2 \left[\sqrt{1-e^2} \left(2 e \mathcal{C}_{0,p}(p e)-\frac{5}{2} \mathcal{C}_{1,p}(p e)+\frac{1}{2} \mathcal{C}_{3,p}(p e)\right)+\left(\frac{7}{2}-2 e^2\right) \mathcal{S}_{1,p}(p e)-\frac{1}{2} \mathcal{S}_{3,p}(p e)\right]\Bigg\} \label{eq:Np:33} \, , \\
N^{4 0}_p =& \frac{1}{14 \sqrt{2}} y^2 (1-3 \nu ) \left(1-e^2\right) e \left(\mathcal{C}_{1,p}(p e)+\frac{5}{6} p \mathcal{S}_{1,p}(p e)\right)\label{eq:Np:40} \, , \\
N^{4 2}_p =& \frac{\sqrt{5}}{42} y^2 (1-3 \nu ) \left(1-e^2\right) \Bigg\{e \mathcal{C}_{1,p}(p e)-p \left[\sqrt{1-e^2} \left(\frac{1}{2} \mathcal{C}_{0,p}(p e)-\mathcal{C}_{2,p}(p e)\right)-\frac{7}{6} e \mathcal{S}_{1,p}(p e)+\mathcal{S}_{2,p}(p e)\right]\nonumber\\
& + \frac{p^2}{3} \sqrt{1-e^2} \left[\sqrt{1-e^2} (\mathcal{C}_{0,p}(p e)-\mathcal{C}_{2,p}(p e))-(2 e \mathcal{S}_{1,p}(p e)-\mathcal{S}_{2,p}(p e))\right]\Bigg\} \label{eq:Np:42} \, , \\
N^{4 4}_p =& \frac{1}{12} \sqrt{\frac{5}{7}} y^2 (1-3 \nu ) \left(1-e^2\right) \Bigg\{ e \mathcal{C}_{1,p}(p e)-p \left[\sqrt{1-e^2} (\mathcal{C}_{0,p}(p e)-2 \mathcal{C}_{2,p}(p e))-\frac{5}{6} e \mathcal{S}_{1,p}(p e)\right]\nonumber\\
& + \frac{p^2}{3} \sqrt{1-e^2} \left[2 \sqrt{1-e^2} (\mathcal{C}_{0,p}(p e)-3 \mathcal{C}_{2,p}(p e))-(10 e \mathcal{S}_{1,p}(p e)-11 \mathcal{S}_{2,p}(p e))\right]\nonumber\\
& - \frac{p^3}{6} \sqrt{1-e^2} \left[\left(15-8 e^2\right) \mathcal{C}_{0,p}(p e)-8 \mathcal{C}_{2,p}(p e)+\mathcal{C}_{4,p}(p e)-\sqrt{1-e^2} (8 e \mathcal{S}_{1,p}(p e)-6 \mathcal{S}_{2,p}(p e)+\mathcal{S}_{4,p}(p e))\right]\Bigg\}\label{eq:Np:44} \, ,
\end{align}
\end{subequations}

\noindent where in $N^{22}_p$ we have introduced $\tilde{f}_{\beta,p}(e)$ as the Fourier series coefficients of $f_\beta(u,e)$, defined in Eq.~\eqref{eq:fbeta}, i.e.

\begin{align}
\tilde{f}_{\beta,p}(e) =& \frac{3 \beta }{\left(1+\beta ^2\right)^2} \Bigg\{ \frac{\beta ^3 \left(53+4 \beta ^2-\beta ^4\right)}{1+\beta^2} J_{p}(p e) + \frac{6-\frac{41 \beta ^2}{3}-\frac{56 \beta ^4}{3}}{1+\beta ^2} J_{p+1}(p e) - \frac{6+13 \beta ^2+7 \beta^4+\frac{2 \beta ^6}{3}-\frac{\beta ^8}{3}}{1+\beta ^2} J_{p-1}(p e) \nonumber\\
& - \frac{\beta^3}{6} J_{p+2}(p e)-\beta  \left(8-6 \beta ^2+\frac{8 \beta^4}{3}-\frac{\beta ^6}{2}\right) J_{p-2}(p e)+\frac{\beta^4}{1+\beta ^2} J_{p+3}(p e) +\frac{1}{1+\beta ^2} J_{p-3}(p e) \nonumber\\
& + \sum_{n=3}^\infty \beta ^{n-3} \left[\beta^4\frac{24}{4-5 n^2+n^4} J_{p+n}(p e) +\left((1 - \beta^2)^4+\frac{2}{n-2}-\frac{4 \beta^2}{n-1}+\frac{4 \beta ^6}{n+1}-\frac{2 \beta ^8}{n+2}\right) J_{p-n}(p e)\right] \Bigg\} \, .
\label{eq:fbetafourier}
\end{align}

This expression for $\tilde{f}_{\beta,p}(e)$ contains infinite sums of Bessel functions of the form

\begin{equation}
    \sum_{n = n_0}^\infty a_n \beta^n J_{p \pm n}(p e) \, .
    \label{eq:Raw_SumOfBesselFuncs}
\end{equation}

While these sums are expected to rapidly converge, we can further speed up their convergence by using the well known recurrence relation of the Bessel functions

\begin{equation}
    \frac{2 \alpha}{z} J_\alpha (z) = J_{\alpha - 1} (z) + J_{\alpha + 1}(z) \, .
    \label{eq:BesselRecurrence}
\end{equation}

Using that $e = 2 \beta/(1 + \beta^2)$, and doing some manipulation, we can write this recurrence relation as

\begin{equation}
    J_{p + n}(p e) = - \frac{p}{n} \left\{J_{p + n}(p e) - \frac{\beta}{1 + \beta^2} \Big[J_{p + n - 1}(p e) + J_{p + n + 1}(p e) \Big]  \right\} \, .
    \label{eq:BesselRecurrenceBeta_p_n}
\end{equation}

Substituting this relation in the sum of Eq.~\eqref{eq:Raw_SumOfBesselFuncs} and appropriately shifting the indices, we find that

\begin{align}
    \sum_{n = n_0}^\infty a_n \beta^n J_{p \pm n}(p e) = & \pm p \Bigg\{\frac{a_{n_0}}{n_0} \frac{\beta^{n_0 + 1}}{1+\beta^2} J_{p \pm (n_0 -1)} (p e) - \left(\frac{a_{n_0}}{n_0} - \frac{a_{n_0+1}}{n_0+1} \frac{\beta^2}{1+\beta^2} \right) \beta^{n_0} J_{p \pm n_0} (p e) \nonumber \\
    & \qquad + \frac{1}{1+\beta^2} \sum_{n=n_0+1}^\infty \left[\left(\frac{a_{n - 1}}{n - 1} - \frac{a_{n}}{n} \right) - \left(\frac{a_{n}}{n} - \frac{a_{n + 1}}{n + 1} \right) \beta^2  \right] \beta^n J_{p \pm n}(p e)  \Bigg\}  \, .
    \label{eq:Simplify_SumOfBesselFuncs}
\end{align}

We note that if $a_n = (c_1 + c_2 \beta^{-2 n}) n$, with $c_1$ and $c_2$ arbitrary constants, the coefficients of the transformed sum vanish, obtaining a closed form expression for the original sum. This is not the case for the sums appearing in Eq.~\eqref{eq:fbetafourier} for $\tilde{f}_{\beta,p}(e)$, where we have coefficients in the sums such that

\begin{align}
    a_n \xrightarrow[n \to \infty]{} \frac{1}{n^k} \implies \left[\left(\frac{a_{n - 1}}{n - 1} - \frac{a_{n}}{n} \right) - \left(\frac{a_{n}}{n} - \frac{a_{n + 1}}{n + 1} \right) \beta^2 \right] \xrightarrow[n \to \infty]{} \frac{k + 1}{n^{k+2}} \left[(1 - \beta^2) + \frac{k + 2}{n} \right] \, ,
\end{align}

\noindent which decays between $1/n^2$ and $1/n^3$ faster than Eq.~\eqref{eq:Raw_SumOfBesselFuncs}. As an example use of the transformation of Eq.~\eqref{eq:Simplify_SumOfBesselFuncs}, we use it to speed up the convergence of the terms that more slowly decay in the sum of Eq.~\eqref{eq:fbetafourier}. Noting that 

\begin{align}
(1 - \beta^2)^4+\frac{2}{n-2}-\frac{4 \beta^2}{n-1}+\frac{4 \beta ^6}{n+1}-\frac{2 \beta ^8}{n+2} =& \frac{24}{4-5 n^2+n^4}+\frac{24 (1 - \beta^2)}{-2 -n + 2 n^2+n^3} + \frac{12 (1 - \beta^2)^2}{2+3 n+n^2} \nonumber\\ 
& + \frac{4 n (1 - \beta^2)^3}{2+3 n+n^2}+\frac{n (1 - \beta^2)^4}{2+n} \, ,
\label{eq:raw_pos_term_fbetafourier}
\end{align}

\noindent and applying Eq.~\eqref{eq:Simplify_SumOfBesselFuncs} to the last two terms, we obtain the following more rapidly converging expression for $\tilde{f}_{\beta,p}(e)$

\begin{align}
\tilde{f}_{\beta,p}(e) =& \frac{3 \beta }{\left(1+\beta ^2\right)^2} \Bigg\{ \frac{\beta ^3 \left(53+4 \beta ^2-\beta ^4\right)}{1+\beta^2} J_{p}(p e) + \frac{6-\frac{41 \beta ^2}{3}-\frac{56 \beta ^4}{3}}{1+\beta ^2} J_{p+1}(p e) - \frac{6+13 \beta ^2+7 \beta^4+\frac{2 \beta ^6}{3}-\frac{\beta ^8}{3}}{1+\beta ^2} J_{p-1}(p e) \nonumber\\
& - \frac{\beta^3}{6} J_{p+2}(p e)-\beta  \left(8-6 \beta ^2+\frac{8 \beta^4}{3}-\frac{\beta ^6}{2}\right) J_{p-2}(p e)+\frac{\beta^4}{1+\beta ^2} J_{p+3}(p e) +\frac{1}{1+\beta ^2} J_{p-3}(p e)  \nonumber\\
& + p \frac{(1 - \beta^2)^3}{1+\beta ^2} \left[\left(\frac{7}{12}-\frac{\beta ^2}{4}\right) J_{p-3}(p e) - \left(\frac{2}{5}-\frac{\beta ^2}{5}\right) \beta J_{p-2}(p e)\right] \nonumber\\
& + \sum_{n=3}^\infty \beta ^{n-3} \Bigg[\beta^4\frac{24}{4-5 n^2+n^4} J_{p+n}(p e) +\bigg\{\frac{24}{4-5 n^2+n^4}+\frac{24 (1 - \beta^2)}{-2 -n + 2 n^2+n^3} + \frac{12 (1 - \beta^2)^2}{2+3 n+n^2} \nonumber\\
&- p \frac{(1 - \beta^2)^3}{1+\beta^2} \frac{1}{(n+2) (n+3)} \left[\frac{1}{n+1}\left(\frac{24}{n} + 10 (1 - \beta^2)\right)+(1 - \beta^2)^2\right] \bigg\} J_{p-n}(p e)\Bigg] \Bigg\} \, .
\label{eq:fbetafourier_enhanced}
\end{align}

\section{Moments of the Fourier Mode Distributions}
\label{sec:appendix:Moments}

In this subsection, we use Eq.~\eqref{eq:Mlm_n} to compute the $n = \{0, 1, 2\}$ moments of the Fourier mode distribution $f_p^{lm}$ (defined in Eq.~\eqref{eq:p_k}). A consistent PN expansion would require computing these moments only to the same relative PN order as used for $F^{lm}$. That is, $\ord{y^2}$ for the $(2,2)$ and $(2,0)$ modes, $\ord{y^1}$ for the $(2,1)$ mode, and $\ord{y^0}$ for all others.  However, to ensure that the resulting expressions remain positive definite, we retain all terms that arise in computing $|F^{lm}|^2$. Nonetheless, in the $(2,2)$, $(2,1)$, and $(2,0)$ modes, the $\ord{y^3}$ and $\ord{y^4}$ terms are incomplete, as we neglect the effects of the 1.5PN and 2PN corrections to $F^{lm}$. To simplify the calculations, we recall from Eq.~\eqref{eq:Flm_def} that $F^{lm}$ can be written as

\begin{equation}
    F^{l m} = \rme^{-\rmi m \phi_F } K^{l m} \, ,
    \label{eq:Flm_phiF}
\end{equation}

\noindent where

\begin{align}
    & \phi_F = W_\phi - \ell = v + 3 y^2 (v - \ell) + \ord{y^3} \nonumber \\
    & = 2 \arctan\left(\sqrt{\frac{1+e}{1-e}} \tan{\frac{u}{2}}\right)+ y^2 \left\{3 \left[2 \arctan\left(\sqrt{\frac{1+e}{1-e}} \tan{\frac{u}{2}}\right)- u + e \sin (u)\right]+\frac{e \sqrt{1-e^2} (4-\nu ) \sin (u)}{1-e \cos (u)}\right\} + \ord{y^3} .
    \label{eq:phiF_def}
\end{align}

While in Eq.~\eqref{eq:expimvpkvml} we Taylor expanded $\rme^{-\rmi m \phi_F }$ to 1PN order, keeping this exponential simplifies the formulas for the moments

\begin{subequations}
\label{eq:Mlm_K}
\begin{align}
M^{l m}_0 & = \Vert \hat{H}^{l m} \Vert^2 = \int_{-\pi}^\pi \frac{\d \ell}{2\pi} \left|F^{l m} \right|^2 = \int_{-\pi}^\pi \frac{\d \ell}{2\pi} \left|K^{l m} \right|^2 \, , \label{eq:Mlm_K:0} \\
M^{l m}_1 & = \mathrm{Im}\left\{ \int_{-\pi}^\pi \frac{\d \ell}{2\pi} F^{l m} \left(\frac{\d F^{l m}}{\d \ell} \right)^{*} \right\} = \int_{-\pi}^\pi \frac{\d \ell}{2\pi}\left\{ \mathrm{Im}\left[ K^{l m} \left(\frac{\d K^{l m}}{\d \ell} \right)^{*} \right] + m \left|K^{l m} \right|^2 \frac{\d \phi_F}{\d \ell} \right\}  \, , \label{eq:Mlm_K:1} \\
M^{l m}_2 & =  \int_{-\pi}^\pi \frac{\d \ell}{2\pi} \left|\frac{\d F^{l m}}{\d \ell} \right|^2 = \int_{-\pi}^\pi \frac{\d \ell}{2\pi} \left|\frac{\d K^{l m}}{\d \ell} - \rmi m K^{l m} \frac{\d \phi_F}{\d \ell} \right|^2 \, . \label{eq:Mlm_K:2}
\end{align}
\end{subequations}

Note that these expressions yield a different $\ord{y^4}$ term in the $(2,2)$ mode compared to what would be obtained by using the 1PN expression for $F^{lm}$ from Eq.~\eqref{eq:F}, or equivalently, by inserting the 1PN $N_p^{lm}$ from Eq.~\eqref{eq:Np} into Eq.~\eqref{eq:Mlm_n_def}. This discrepancy is not relevant when analyzing the general behavior of the modes. Nonetheless, in certain cases, such as when computing the Fourier modes that have to be included (see Sec.~\ref{sec:Properties:OptimumInclusion}), it is necessary to use the exact value of the norm that would be obtained by summing $|N_p^{lm}|^2$ using the $N_p^{lm}$ from Eq.~\eqref{eq:Np}. This will be discussed further in App.~\ref{sec:appendix:Moments:Norms}.

\subsection{Norms}
\label{sec:appendix:Moments:Norms}

Substituting the 1PN $K^{l m}$ in Eq.~\eqref{eq:Mlm_K:0} and integrating, we obtain the following norms of the GW modes

\begin{subequations}
\label{eq:ezNorms}
\begin{align}
\Vert \hat{H}^{2 0} \Vert^2 =& \frac{2}{3} \frac{e^2}{\left(1+\sqrt{1-e^2}\right) \sqrt{1-e^2}} + y^2 e^2 \left\{\frac{38}{21}-\frac{34 \nu }{63}+\frac{1}{\sqrt{1-e^2}}\left[-\frac{30}{7}+\frac{40 \nu}{63}+\frac{1}{1+\sqrt{1-e^2}} \left(-\frac{6}{7}-\frac{34 \nu}{63}\right) \right]\right\} \nonumber \\
& + y^4 e^2 \Bigg\{\frac{2 (27+17\nu )^2}{5292 \left(1+\sqrt{1-e^2}\right) \sqrt{1-e^2}}-\frac{57}{49}-\frac{170 \nu }{441}+\frac{289 \nu^2}{1323}+\left(-\frac{361}{294}+\frac{323 \nu }{441}-\frac{289 \nu^2}{2646}\right) e^2 \nonumber\\
& + \frac{1}{\sqrt{1-e^2}} \left[\frac{1081}{147}+\frac{143 \nu}{147}-\frac{331 \nu ^2}{1323}+\left(\frac{2363}{294}-\frac{313 \nu}{147}+\frac{809 \nu ^2}{5292}\right) e^2\right] \Bigg\} \, , \label{eq:ezNorms:20} \\
\Vert \hat{H}^{2 1} \Vert^2 =& y^2 \left(\delta \mu - \frac{3}{2} y \delta\chi \right)^2 \frac{2}{9} \frac{2+e^2}{\sqrt{1-e^2}}  \, , \label{eq:ezNorms:21} \\
\Vert \hat{H}^{2 2} \Vert^2 =& \left\{\frac{5}{\sqrt{1-e^2}}-1\right\} + y^2 \left\{\frac{9}{7}+\frac{17 \nu}{21}+\left(\frac{19}{7}-\frac{17 \nu }{21}\right) e^2+\frac{1}{\sqrt{1-e^2}} \left[-\frac{65}{3}+\frac{29 \nu }{3}+\left(\frac{5}{3}+\frac{16 \nu}{3}\right) e^2\right] \right\} \nonumber \\
& + y^4 \Bigg\{-\frac{81}{196}-\frac{51\nu }{98}-\frac{289 \nu ^2}{1764}+\left(-\frac{171}{98}-\frac{85 \nu}{147}+\frac{289 \nu ^2}{882}\right) e^2+\left(-\frac{361}{196}+\frac{323\nu }{294}-\frac{289 \nu ^2}{1764}\right) e^4 \nonumber\\
& + \frac{1}{\sqrt{1-e^2}} \bigg[\frac{46525}{1764}-\frac{23081 \nu }{882}+\frac{12389 \nu^2}{1764}+\left(\frac{2774}{441}-\frac{8375 \nu }{882}+\frac{5377 \nu^2}{882}\right) e^2 \nonumber\\
& + \left(-\frac{14057}{3528}-\frac{1513 \nu}{441}+\frac{5437 \nu ^2}{3528}\right) e^4\bigg] \Bigg\} \, , \label{eq:ezNorms:22} \\
\Vert \hat{H}^{3 0} \Vert^2 =& y^4 (1-3 \nu )^2 \frac{e^2}{\sqrt{1-e^2}} \left(\frac{1}{84}+\frac{e^2}{336}\right) \, , \label{eq:ezNorms:30} \\
\Vert \hat{H}^{3 1} \Vert^2 =& y^2 \delta \mu^2 \left\{\frac{1}{14} \left(1-e^2\right)+\frac{1}{\sqrt{1-e^2}} \left[-\frac{5}{72}+\frac{145 e^2}{1008}\right]\right\}  \, , \label{eq:ezNorms:31} \\
\Vert \hat{H}^{3 2} \Vert^2 =& y^4 (1-3 \nu )^2 \frac{1}{\sqrt{1-e^2}} \left(\frac{20}{63}+\frac{485 e^2}{504}+\frac{35 e^4}{288}\right) \, , \label{eq:ezNorms:32} \\
\Vert \hat{H}^{3 3} \Vert^2 =& y^2 \delta \mu^2  \left\{\frac{5}{42} \left(1-e^2\right)+\frac{1}{\sqrt{1-e^2}} \left[\frac{55}{24}+\frac{115 e^2}{48}\right] \right\} \, , \label{eq:ezNorms:33} \\
\Vert \hat{H}^{4 0} \Vert^2 =& y^4 (1-3 \nu )^2 \left\{\frac{1}{\sqrt{1-e^2}} \left[\frac{1}{98}-\frac{179 e^2}{7056}+\frac{67 e^4}{3136}\right] -\frac{1}{98} \left(1-e^2\right)^2\right\} \, , \label{eq:ezNorms:40} \\
\Vert \hat{H}^{4 2} \Vert^2 =& y^4 (1-3 \nu )^2 \left\{\frac{1}{\sqrt{1-e^2}} \left[\frac{65}{3969}-\frac{485 e^2}{15876}+\frac{25 e^4}{392}\right] -\frac{5}{441} \left(1-e^2\right)^2\right\} \, , \label{eq:ezNorms:42} \\
\Vert \hat{H}^{4 4} \Vert^2 =& y^4 (1-3 \nu )^2 \left\{\frac{1}{\sqrt{1-e^2}} \left[\frac{5165}{2268}+\frac{119765 e^2}{18144}+\frac{1035 e^4}{896}\right] -\frac{5}{252} \left(1-e^2\right)^2\right\} \, . \label{eq:ezNorms:44}
\end{align}
\end{subequations}

As previously mentioned, the $\ord{y^4}$ term of the $(l,m)=(2,2)$ norm does not correspond to what would be obtained if we substituted the Fourier mode coefficients $N^{l m}_p$ of Eq.~\eqref{eq:Np} in Eq.~\eqref{eq:normHlm}, since we have not consistently PN expanded the exponentials in Eq.~\eqref{eq:expimvpkvml} and Eq.~\eqref{eq:Flm_phiF}. To obtain a norm consistent with $N^{l m}_p$, we use that in that case

\begin{equation}
    F^{2 2} = \rme^{-\rmi 2 \phi_{F,0\mathrm{PN}} } \left[ (1 - \rmi y^2 2 \phi_{F,1\mathrm{PN}}) K^{2 2}_{0 \mathrm{PN}} + y^2 K^{2 2}_{1 \mathrm{PN}} \right] \, .
    \label{eq:F22_PN_consistent}
\end{equation}

Substituting this in Eq.~\eqref{eq:normHlm_Flm} and expanding we obtain

\begin{equation}
    \Vert \hat{H}^{2 2} \Vert^2 = \int_{-\pi}^{\pi}  \frac{\d \ell}{2 \pi} |F^{2 2}|^2 = \int_{-\pi}^{\pi}  \frac{\d \ell}{2 \pi} \left\{ \left|K^{2 2}_{0 \mathrm{PN}} + y^2 K^{2 2}_{1 \mathrm{PN}} \right|^2 + y^4 2 \phi_{F,1\mathrm{PN}} \left[ 2 \phi_{F,1\mathrm{PN}}  \left|K^{2 2}_{0 \mathrm{PN}}\right|^2 + 2 \mathrm{Im}\left\{K^{2 2}_{0 \mathrm{PN}} (K^{2 2}_{1 \mathrm{PN}})^{*} \right\} \right] \right\}  \, .
    \label{eq:normH22_F22_PN_consistent}
\end{equation}

While the first term correspond to the relatively simple integral computed in Eq.~\eqref{eq:ezNorms:22}, the second term, is much harder to integrate, due to the complicated expression of $\phi_F$ (Eq.~\eqref{eq:phiF_def}). Nonetheless, the result can be computed analytically, yielding the following PN consistent norm of the $(2,2)$ mode:

\begin{align}
\Vert \hat{H}^{2 2} \Vert^2 =& \left\{\frac{5}{\sqrt{1-e^2}}-1\right\} + y^2 \left\{\frac{9}{7}+\frac{17 \nu}{21}+\left(\frac{19}{7}-\frac{17 \nu }{21}\right) e^2+\frac{1}{\sqrt{1-e^2}} \left[-\frac{65}{3}+\frac{29 \nu }{3}+\left(\frac{5}{3}+\frac{16 \nu}{3}\right) e^2\right] \right\} \nonumber \\
& + y^4 \Bigg\{ \frac{72999}{196}-\frac{2851 \nu }{98}+\frac{6767 \nu ^2}{1764}+\left(\frac{5625}{98}+\frac{4115 \nu }{147}-\frac{3239 \nu ^2}{882}\right) e^2+\left(-\frac{361}{196}+\frac{323 \nu }{294}-\frac{289 \nu ^2}{1764}\right) e^4 \nonumber \\
& +\frac{1}{\sqrt{1-e^2}}\Bigg[-\frac{611195}{1764}+\frac{2119 \nu }{882}+\frac{5333 \nu ^2}{1764}+\left(\frac{223463}{441}-\frac{117743 \nu }{882}+\frac{11677 \nu ^2}{882}\right) e^2 \nonumber\\
& +\left(-\frac{141737}{3528}+\frac{11234 \nu }{441}-\frac{5147 \nu ^2}{3528}\right) e^4\Bigg]+48 \left[15-(4-\nu ) \sqrt{1-e^2}\right] \log \left(\frac{1+\sqrt{1-e^2}}{2 \sqrt{1-e^2}}\right) \nonumber \\
& + 72 \left[\frac{5}{\sqrt{1-e^2}}-1\right] \mathrm{Li}_2 \left[\frac{e^2}{\left(1+\sqrt{1-e^2}\right)^2} \right] \Bigg\} \, ,
\label{eq:normH22}
\end{align}

\noindent where $\mathrm{Li}_2 (z)$ is the dilogarithm (or Spence's function), defined as

\begin{equation}
    \mathrm{Li}_2 (z) = - \int_{0}^z \frac{\log{(1 - u)}}{u} \d u = \sum_{k = 1}^\infty \frac{z^k}{k^2} \, . 
    \label{eq:polylog_def}
\end{equation}

\subsection{First unnormalized moments}
\label{sec:appendix:Moments:First}

Substituting the 1PN $K^{l m}$ and $\phi_F$ in Eq.~\eqref{eq:Mlm_K:1} and operating, we obtain the following first unnormalized moments of the GW modes, that are closely related to their mean frequency,

\begin{subequations}
\label{eq:ezM1}
\begin{align}
M_1^{2 0} =& 0 \, , \label{eq:ezM1:20} \\
M_1^{2 1} =& \frac{y^2 \left(\delta \mu - \frac{3}{2} y \delta\chi \right)^2}{\left(1-e^2\right)^2} \left(\frac{4}{9}+\frac{4 e^2}{3}+\frac{e^4}{6}\right) \, , \label{eq:ezM1:21} \\
M_1^{2 2} =& \frac{1}{\left(1-e^2\right)^2}\Bigg\{ 8 + 7 e^2 + y^2 \Bigg[-\frac{226}{21}+\frac{440 \nu }{21}+\left(\frac{991}{21}+\frac{485\nu }{21}\right) e^2+\left(\frac{2155}{84}-\frac{65 \nu }{42}\right) e^4-30 \left(1-e^2\right)^{3/2}\Bigg] \nonumber\\
& + y^4 \Bigg[-\frac{5767}{441}-\frac{10751 \nu }{441}+\frac{6050 \nu^2}{441}+\left(-\frac{28703}{252}+\frac{10061 \nu }{126}+\frac{613 \nu^2}{28}\right) e^2 +\left(-\frac{2413}{392}+\frac{15297 \nu}{392}-\frac{295 \nu ^2}{588}\right) e^4 \nonumber \\
& + \left(-\frac{17845}{588}+\frac{23353 \nu }{1764}-\frac{3853 \nu^2}{3528}\right) e^6+\left(1-e^2\right)^{3/2} \Big\{65-29 \nu +(-5-16 \nu) e^2\Big\}\Bigg] \Bigg\} \, , \label{eq:ezM1:22} \\
M_1^{3 0} =& 0 \, , \label{eq:ezM1:30} \\
M_1^{3 1} =& \frac{y^2 \delta \mu ^2}{\left(1-e^2\right)^2} \left(\frac{1}{504}+\frac{11 e^2}{168}+\frac{67 e^4}{448}\right) \, , \label{eq:ezM1:31} \\
M_1^{3 2} =& \frac{y^4 (1-3 \nu )^2}{\left(1-e^2\right)^2} \left( \frac{40}{63}+\frac{1145 e^2}{252}+\frac{545 e^4}{168}+\frac{115 e^6}{672} \right) \, , \label{eq:ezM1:32} \\
M_1^{3 3} =& \frac{y^2 \delta \mu ^2}{\left(1-e^2\right)^2} \left(\frac{405}{56}+\frac{3695 e^2}{168}+\frac{4645 e^4}{1344}\right) \, , \label{eq:ezM1:33} \\
M_1^{4 0} =& 0 \, , \label{eq:ezM1:40} \\
M_1^{4 2} =& \frac{y^4 (1-3 \nu )^2}{\left(1-e^2\right)^2} \left(\frac{40}{3969}-\frac{5 e^2}{252}+\frac{2725 e^4}{10584}+\frac{1385 e^6}{14112}\right) \, , \label{eq:ezM1:42} \\
M_1^{4 4} =& \frac{y^4 (1-3 \nu )^2}{\left(1-e^2\right)^2} \left(\frac{5120}{567}+\frac{22205 e^2}{378}+\frac{29125 e^4}{756}+\frac{1045 e^6}{504}\right)\, . \label{eq:ezM1:44}
\end{align}
\end{subequations}

\subsection{Second unnormalized moments}
\label{sec:appendix:Moments:Second}

Substituting the 1PN $K^{l m}$ and $\phi_F$ in Eq.~\eqref{eq:Mlm_K:2} and operating, we obtain the following second unnormalized moments of the GW modes, that are closely related to their frequency spread,

\begin{subequations}
\label{eq:ezM2}
\begin{align}
M_2^{2 0} =& \frac{e^2}{\left(1-e^2\right)^{7/2}} \Bigg\{\frac{1}{3}+\frac{e^2}{12} + y^2 \left[-\frac{61}{21}-\frac{11 \nu }{63}+\left(-\frac{709}{84}+\frac{41 \nu }{84}\right) e^2+\left(-\frac{97}{84}+\frac{13 \nu }{126}\right) e^4\right]\nonumber\\
& + y^4\Bigg[\frac{3721}{588}+\frac{671 \nu }{882}+\frac{121 \nu^2}{5292}+\left(\frac{126041}{2352}-\frac{1789 \nu }{1176}-\frac{2251 \nu^2}{21168}\right) e^2 +\left(\frac{22285}{392}-\frac{8059 \nu}{1176}+\frac{2071 \nu ^2}{10584}\right) e^4 \nonumber \\
& +\left(\frac{5465}{1176}-\frac{5395 \nu }{7056}+\frac{2785 \nu^2}{84672}\right) e^6\Bigg]\Bigg\} \, , \label{eq:ezM2:20} \\
M_2^{2 1} =& \frac{y^2 \left(\delta \mu - \frac{3}{2} y \delta\chi \right)^2}{\left(1-e^2\right)^{7/2}} \left(\frac{4}{9}+\frac{38 e^2}{9}+\frac{23 e^4}{6}+\frac{e^6}{4}\right) \, , \label{eq:ezM2:21} \\
M_2^{2 2} =& \frac{1}{\left(1-e^2\right)^{7/2}} \Bigg\{ 16 + \frac{97 e^2}{2}+\frac{49 e^4}{8}+y^2 \Bigg[\frac{304}{21}+\frac{880 \nu }{21}+\left(\frac{16453}{42}+\frac{5213 \nu }{42}\right) e^2+\left(\frac{19939}{56}-\frac{23 \nu }{56}\right) e^4 \nonumber \\ 
& +\left(\frac{825}{56}-\frac{209 \nu }{84}\right) e^6-12 \left(8+7 e^2\right) \left(1-e^2\right)^{3/2}\Bigg] + y^4 \Bigg[\frac{80824}{441}+\frac{8360 \nu }{441}+\frac{12100 \nu ^2}{441} \nonumber \\
& + \left(\frac{1885}{3528}+\frac{750569 \nu}{1764}+\frac{104915 \nu ^2}{1176}\right) e^2+\left(\frac{34091653}{14112}+\frac{1393739 \nu }{7056}+\frac{152581 \nu ^2}{14112}\right) e^4 \nonumber \\
& +\left(\frac{807595}{1764}-\frac{283519 \nu }{2352}+\frac{46639 \nu ^2}{7056}\right) e^6+\left(\frac{84159}{6272}-\frac{267 \nu }{49}+\frac{34873 \nu^2}{56448}\right) e^8 \nonumber \\
& -\left(1-e^2\right)^{3/2} \left\{\frac{556}{7}+\frac{880 \nu }{7}+\left(\frac{5636}{7}+\frac{466 \nu }{7}\right)  e^2+\left(\frac{4339}{14}-\frac{275 \nu }{7}\right) e^4\right\}\Bigg] \Bigg\} \, , \label{eq:ezM2:22} \\
M_2^{3 0} =& \frac{y^4 (1-3 \nu )^2}{\left(1-e^2\right)^{7/2}} e^2 \left(\frac{1}{84}+\frac{37 e^2}{336}+\frac{59 e^4}{672}+\frac{9 e^6}{1792}\right) \, , \label{eq:ezM2:30} \\
M_2^{3 1} =& \frac{y^2 \delta \mu ^2}{\left(1-e^2\right)^{7/2}} \left(\frac{1}{504}+\frac{127 e^2}{1008}+\frac{995 e^4}{1344}+\frac{97 e^6}{896}\right) \, , \label{eq:ezM2:31} \\
M_2^{3 2} =& \frac{y^4 (1-3 \nu )^2}{\left(1-e^2\right)^{7/2}} \left(\frac{80}{63}+\frac{1295 e^2}{72}+\frac{68885 e^4}{2016}+\frac{46595 e^6}{4032}+\frac{1315 e^8}{3584}\right)\, , \label{eq:ezM2:32} \\
M_2^{3 3} =& \frac{y^2 \delta \mu ^2}{\left(1-e^2\right)^{7/2}} \left(\frac{1215}{56}+\frac{7045 e^2}{48}+\frac{44375 e^4}{448}+\frac{13285 e^6}{2688}\right)\, , \label{eq:ezM2:33} \\
M_2^{4 0} =& \frac{y^4 (1-3 \nu )^2}{\left(1-e^2\right)^{7/2}} e^2 \left(\frac{1}{7056}+\frac{27 e^2}{3136}+\frac{3379 e^4}{56448}+\frac{349 e^6}{50176}\right) \, , \label{eq:ezM2:40} \\
M_2^{4 2} =& \frac{y^4 (1-3 \nu )^2}{\left(1-e^2\right)^{7/2}} \left(\frac{80}{3969}-\frac{1115 e^2}{15876}+\frac{1625 e^4}{1323}+\frac{67465 e^6}{42336}+\frac{3145 e^8}{28224}\right) \, , \label{eq:ezM2:42} \\
M_2^{4 4} =& \frac{y^4 (1-3 \nu )^2}{\left(1-e^2\right)^{7/2}} \left(\frac{20480}{567}+\frac{1102895 e^2}{2592}+\frac{16004785 e^4}{24192}+\frac{8704825 e^6}{48384}+\frac{581405 e^8}{129024}\right) \, . \label{eq:ezM2:44}
\end{align}
\end{subequations}

\section{Asymptotic Expansion of Fourier Mode Amplitudes}
\label{sec:appendix:Asymptotic}

In this section we study how the expressions for $N^{l m}_p$ behave as $p \to \pm \infty$. To this end we use Debye's asymptotic expansion of Bessel functions, given by~\cite{Abramowitz_and_Stegun}

\begin{subequations}
\label{eq:DebyeAssymp}
\begin{align}
    J_p (p e) =& \left\{1 + \sum_{k=1}^\infty \frac{1}{|p|^k} u_k \left( \frac{1}{\sqrt{1-e^2}} \right) \right\} D_{|p|}^\mathrm{LO}(e) \, , \\
    J'_p (p e) =& \mathrm{sign}(p) \frac{\sqrt{1-e^2}}{e} \left\{1 + \sum_{k=1}^\infty \frac{1}{|p|^k} v_k \left( \frac{1}{\sqrt{1-e^2}} \right) \right\}D_{|p|}^\mathrm{LO}(e)  \, , \\   
    D_{|p|}^\mathrm{LO}(e) =& \frac{1}{\sqrt{2 \pi |p| \sqrt{1-e^2}}} \exp\left\{ - |p| \left[ \log\left( \frac{1 + \sqrt{1-e^2}}{e} \right)  - \sqrt{1-e^2}\right] \right\} \, , \\
    u_1(t) =& \frac{3 t - 5 t^3}{24} \, , \\
    u_2(t) =& \frac{81 t^2-462 t^4+385 t^6}{1152} \, , \\
    u_3(t) =& \frac{30375 t^3-369603 t^5+765765 t^7-425425 t^9}{414720} \, , \\
    u_4(t) =& \frac{4465125 t^4-94121676 t^6+349922430 t^8-446185740 t^{10}+185910725 t^{12}}{39813120} \, , \\
    v_1(t) =& \frac{-9 t+7 t^3}{24} \, , \\
    v_2(t) =& \frac{-135 t^2+594 t^4-455 t^6}{1152} \, , \\
    v_3(t) =& \frac{-42525 t^3+451737 t^5-883575 t^7+475475 t^9}{414720} \, , \\
    v_4(t) =& \frac{-5740875 t^4+111234708 t^6-396578754 t^8+493152660 t^{10}-202076875 t^{12}}{39813120} \, . 
\end{align}
\end{subequations}

In order to use this expansion we need to write $N^{l m}_p$ in terms of $J_p (p e)$ and $J'_p (p e)$. This can be achieved by repeatedly applying to Eq.~\eqref{eq:Np} the recurrence relation of Eq.~\eqref{eq:BesselRecurrence}, and the following relation for the derivative of Bessel functions

\begin{equation}
    2 J'_\alpha (z) = J_{\alpha - 1}(z) - J_{\alpha + 1} (z) \, .
    \label{eq:dJBessel}
\end{equation}

Doing this, we obtain the following expressions, valid for $p \neq 0$ and $e \neq 0$,

\begin{subequations}
\label{eq:NpJpdJp}
\begin{align}
N^{2 0}_p =& \sqrt{\frac{2}{3}} \left\{\left[1-y^2 \left(\frac{9}{14}+\frac{17 \nu }{42}+e^2 \left(\frac{19}{14}-\frac{17 \nu }{42}\right)\right)\right] J_p (p e)-p y^2 \left(1-e^2\right) \left(\frac{26}{7}-\frac{\nu }{7}\right) e J'_p (p e)\right\} \label{eq:NpJpdJp:20} \, ,  \\
N^{2 1}_p =& \frac{2}{3} \rmi p y \left(\delta \mu - \frac{3}{2} y \delta\chi \right) \frac{1-e^2}{e} \left\{ \sqrt{1-e^2} J_p (p e)+e J'_p (p e) \right\} \label{eq:NpJpdJp:21} \, ,  \\
N^{2 2}_p =& \frac{2}{e^2} \left\{\left[-1+\frac{e^2}{2}+\left(1-e^2\right)^{3/2} p\right] J_p (p e)+\sqrt{1-e^2} \left[-1+\sqrt{1-e^2} p\right] e J'_p (p e)\right\} + \frac{y^2}{e^2} \Bigg\{ \Bigg[\frac{37}{7}-\frac{111 e^2}{14}-\frac{19 e^4}{14} \nonumber\\
& - \nu  \left(1-e^2\right) \left(\frac{67}{21}+\frac{17 e^2}{42}\right)+p \sqrt{1-e^2} \left(-\frac{113}{21}-\frac{39 e^2}{7}+\frac{356 e^4}{21}+\nu  \left(1-e^2\right) \left(\frac{73}{21}+\frac{11 e^2}{21}\right)\right) \nonumber \\
& + p^2 \frac{2}{21} \left(1-e^2\right)^3 (1-3 \nu )\Bigg] J_p (p e)+\Bigg[\sqrt{1-e^2} \left(\frac{37}{7} \left(1-e^2\right)-\nu  \left(\frac{67}{21}-\frac{25 e^2}{21}\right)\right) \nonumber \\
& + p \left(-\frac{113}{21}+\frac{262 e^2}{21}-\frac{23 e^4}{21}+\nu  \left(1-e^2\right) \left(\frac{73}{21}+\frac{8 e^2}{21}\right)\right) + p^2 \frac{2}{21} \left(1-e^2\right)^{5/2} (1-3 \nu )\Bigg] e J'_p (p e) \Bigg\} + y^2 p  \tilde{f}_{\beta,p}(e) \label{eq:NpJpdJp:22} \, ,  \\
N^{3 0}_p =& \frac{1}{\sqrt{42}} p y^2 (1-3 \nu ) \left(1-e^2\right)^{3/2} J_p (p e) \label{eq:NpJpdJp:30} \, ,  \\
N^{3 1}_p =& \frac{\rmi}{\sqrt{14}} y \delta \mu \frac{\sqrt{1-e^2}}{e} \left[1-\frac{5}{6} \sqrt{1-e^2} p\right] \left\{\sqrt{1-e^2} J_p (p e)+e J'_p (p e)\right\}\label{eq:NpJpdJp:31} \, ,  \\
N^{3 2}_p =& \frac{1}{3} \sqrt{\frac{5}{7}} p y^2 (1-3 \nu ) \frac{\left(1-e^2\right)^{3/2}}{e^2} \left\{ \left[-1+\frac{e^2}{2}+\left(1-e^2\right)^{3/2} p\right] J_p (p e)+\sqrt{1-e^2} \left[-1+\sqrt{1-e^2} p\right] e J'_p (p e) \right\} \label{eq:NpJpdJp:32} \, ,  \\
N^{3 3}_p =& \rmi \sqrt{\frac{5}{42}} y \delta \mu  \frac{\sqrt{1-e^2}}{e^3} \Bigg\{\left[\sqrt{1-e^2} \left(-4+e^2\right)+\left(6-\frac{5 e^2}{2}\right) \left(1-e^2\right) p-2 \left(1-e^2\right)^{5/2} p^2\right] J_p (p e) \nonumber\\
& + \left[-4+3 e^2+\sqrt{1-e^2} \left(6-\frac{7 e^2}{2}\right) p-2 \left(1-e^2\right)^2 p^2\right] e J'_p (p e)\Bigg\} \label{eq:NpJpdJp:33} \, ,  \\
N^{4 0}_p =& \frac{1}{7 \sqrt{2}} y^2 (1-3 \nu ) \left(1-e^2\right) \left\{J_p (p e)-\frac{5}{6}  p e J'_p (p e) \right\} \label{eq:NpJpdJp:40} \, ,  \\
N^{4 2}_p =& \frac{\sqrt{5}}{21} y^2 (1-3 \nu ) \frac{1-e^2}{e^2} \Bigg\{\left[-2+e^2+\sqrt{1-e^2} \left(\frac{8}{3}-\frac{3 e^2}{2}\right) p-\frac{2}{3} \left(1-e^2\right)^2 p^2\right] J_p (p e) \nonumber \\
& + \left[-2 \sqrt{1-e^2}+\left(\frac{8}{3}-\frac{11 e^2}{6}\right) p-\frac{2}{3} \left(1-e^2\right)^{3/2} p^2\right] e J'_p (p e)\Bigg\} \label{eq:NpJpdJp:42} \, ,  \\
N^{4 4}_p =& \frac{1}{6}\sqrt{\frac{5}{7}} y^2 (1-3 \nu ) \frac{1-e^2}{e^4} \nonumber\\
& \times \Bigg\{ \Bigg[8-8 e^2+e^4+\sqrt{1-e^2} \left(-\frac{44}{3}+\frac{46 e^2}{3}-3 e^4\right) p+\left(8-\frac{8 e^2}{3}\right) \left(1-e^2\right)^2 p^2-\frac{4}{3} \left(1-e^2\right)^{7/2} p^3\Bigg] J_p (p e) \nonumber\\
& +\Bigg[\sqrt{1-e^2} \left(8-4 e^2\right)+\left(-\frac{44}{3}+\frac{56 e^2}{3}-\frac{29 e^4}{6}\right) p+\left(8-\frac{10 e^2}{3}\right) \left(1-e^2\right)^{3/2} p^2-\frac{4}{3} \left(1-e^2\right)^3 p^3\Bigg] e J'_p (p e)\Bigg\} \label{eq:NpJpdJp:44} \, ,
\end{align}
\end{subequations}

\noindent where the same recurrence relations can be applied to $\tilde{f}_{\beta,p}(e)$, appearing in Eq.~\eqref{eq:NpJpdJp:22} and defined in Eq.~\eqref{eq:fbetafourier}, to obtain

\begin{align}
    \tilde{f}_{\beta,p}(e) =& \frac{1}{4 \beta ^2}\Bigg\{\Bigg[\frac{6-48 \beta ^2-167 \beta ^4+299 \beta ^6-136 \beta ^8+14 \beta ^{10}-11 \beta ^{12}+3 \beta ^{14}}{\left(1+\beta ^2\right)^3}-\frac{18-54 \beta ^2-5 \beta ^4+3 \beta ^6-13 \beta ^8+3 \beta ^{10}}{(1+\beta ^2) p} \nonumber \\
    & + \frac{12 \left(1+\beta ^4\right)}{p^2}\Bigg] J_p (p e) + \Bigg[\frac{12-240 \beta ^2+130 \beta ^4+18 \beta ^6-22 \beta ^8+6 \beta ^{10}}{\left(1+\beta ^2\right)^2}-\frac{36-96 \beta ^2+106 \beta ^4-32 \beta ^6+6 \beta ^8}{(1+\beta ^2) p} \nonumber \\
    & + \frac{24 (1-\beta^2)}{p^2}\Bigg] \beta  J'_p (p e) \Bigg\} + \frac{3 \beta }{\left(1+\beta ^2\right)^2} \tilde{f}^\mathrm{sum}_{\beta,p}(e) \, ,
    \label{eq:fbetaJpdJp}
\end{align}

\noindent and $\tilde{f}^\mathrm{sum}_{\beta,p}(e)$ contains the infinite sums of Eq.~\eqref{eq:fbetafourier}, i.e.

\begin{align}
    \tilde{f}^\mathrm{sum}_{\beta,p}(e) =& \sum_{n=3}^\infty \beta ^{n-3} \left[\beta^4\frac{24}{4-5 n^2+n^4} J_{p+n}(p e) +\left((1 - \beta^2)^4+\frac{2}{n-2}-\frac{4 \beta^2}{n-1}+\frac{4 \beta ^6}{n+1}-\frac{2 \beta ^8}{n+2}\right) J_{p-n}(p e)\right] \, .
    \label{eq:fbetasum}
\end{align}

The asymptotic expansions of the expressions in Eq.~\eqref{eq:NpJpdJp} and Eq.~\eqref{eq:fbetaJpdJp} can be easily obtained by substituting Debye's asymptotic expansion of $J_p(p e)$ and $J'_p(p e)$. However, obtaining an asymptotic expansion for Eq.~\eqref{eq:fbetasum} is not so straightforward. To compute how this term behaves as $p \to \pm \infty$ we use that, from Eq.~\eqref{eq:DebyeAssymp},

\begin{align}
    \frac{J_{p + n} (p e)}{J_{p + n_0} (p e)} \xrightarrow[|p| \to \infty]{} \beta^{(n - n_0) \mathrm{sign}(p)} \exp\left\{-\frac{1}{2 |p| \sqrt{ 1 - e^2}} \left[n^2 - n_0^2 + \mathrm{sign}(p) \frac{n - n_0}{\sqrt{1-e^2}} \right]\right\} \left(1 + \ord{\frac{1}{p^2}} \right) \label{eq:BesselRatioDebyeAssymp} \, .
\end{align}

With this, we can show that $\tilde{f}^\mathrm{sum}_{\beta,p}(e)/D_{|p|}^\mathrm{LO} \xrightarrow[p \to \infty]{} \ord{\sqrt{p}}$, and therefore, when $p \to \infty$, $\tilde{f}_{\beta,p}(e)$ grows slower than the $\ord{p^2}$ terms that appear in Eq.~\eqref{eq:NpJpdJp:22}. Meanwhile, when $p \to -\infty$,

\begin{align}
    \frac{\tilde{f}^\mathrm{sum}_{\beta,p}}{D_{|p|}^\mathrm{LO}} \xrightarrow[p \to - \infty]{} & \frac{5 \beta }{6}+3 \beta ^3-4 \beta ^5+\frac{7 \beta ^7}{3}-\frac{\beta ^9}{2} \nonumber\\
    & + \frac{\beta}{(1-\beta^2)^3 p} \left[\frac{377}{72}+\frac{257 \beta ^2}{24}-\frac{221 \beta ^4}{8}+\frac{241 \beta ^6}{72}+\frac{475 \beta ^8}{24}-\frac{97 \beta ^{10}}{8}+\frac{287 \beta ^{12}}{72}-\frac{13 \beta ^{14}}{24}\right] + \ord{\frac{1}{p^{3/2}}} ,
    \label{eq:fbetasumAsympNeg}
\end{align}

\noindent which has to be computed up to $\ord{1/p}$ since the leading $\ord{p^0}$ term cancels when substituting $\tilde{f}^\mathrm{sum}_{\beta,p}$ in Eqs.~(\ref{eq:NpJpdJp:22},\ref{eq:fbetaJpdJp}). With all the results above, we obtain the leading order $p \to \infty$ asymptotic expansion of $N^{l m}_p$,

\begin{subequations}
\label{eq:NpAsympPlus}
\begin{align}
\frac{N^{2 0}_p}{D_{|p|}^\mathrm{LO}} \xrightarrow[p \to \infty]{} & \sqrt{\frac{2}{3}} \left\{1-y^2 \left(\frac{26}{7}-\frac{\nu }{7}\right) \left(1-e^2\right)^{3/2} p\right\}  \label{eq:NpAsympPlus:20} \, , \\
\frac{N^{2 1}_p}{D_{|p|}^\mathrm{LO}} \xrightarrow[p \to \infty]{} & \frac{4}{3} \rmi y \left(\delta \mu - \frac{3}{2} y \delta\chi \right)  \frac{\left(1-e^2\right)^{3/2}}{e} p \label{eq:NpAsympPlus:21} \, , \\
\frac{N^{2 2}_p}{D_{|p|}^\mathrm{LO}} \xrightarrow[p \to \infty]{} &  \frac{1}{e^2}\left\{ 4 \left(1-e^2\right)^{3/2} p + \frac{4}{21} y^2 (1-3 \nu ) \left(1-e^2\right)^3 p^2 \right\} \label{eq:NpAsympPlus:22} \, , \\
\frac{N^{3 0}_p}{D_{|p|}^\mathrm{LO}} \xrightarrow[p \to \infty]{} & \frac{1}{\sqrt{42}} y^2 (1-3 \nu ) \left(1-e^2\right)^{3/2} p \label{eq:NpAsympPlus:30} \, , \\
\frac{N^{3 1}_p}{D_{|p|}^\mathrm{LO}} \xrightarrow[p \to \infty]{} & -\frac{5 \rmi}{3 \sqrt{14}} y \delta \mu  \frac{\left(1-e^2\right)^{3/2}}{e} p \label{eq:NpAsympPlus:31} \, , \\
\frac{N^{3 2}_p}{D_{|p|}^\mathrm{LO}} \xrightarrow[p \to \infty]{} & \frac{2}{3} \sqrt{\frac{5}{7}} y^2 (1-3 \nu ) \frac{\left(1-e^2\right)^3}{e^2} p^2  \label{eq:NpAsympPlus:32} \, , \\
\frac{N^{3 3}_p}{D_{|p|}^\mathrm{LO}} \xrightarrow[p \to \infty]{} & -4 \rmi \sqrt{\frac{5}{42}} y \delta \mu  \frac{\left(1-e^2\right)^3}{e^3} p^2  \label{eq:NpAsympPlus:33} \, , \\
\frac{N^{4 0}_p}{D_{|p|}^\mathrm{LO}} \xrightarrow[p \to \infty]{} & -\frac{5}{42 \sqrt{2}} y^2 (1-3 \nu ) \left(1-e^2\right)^{3/2} p \label{eq:NpAsympPlus:40} \, , \\
\frac{N^{4 2}_p}{D_{|p|}^\mathrm{LO}} \xrightarrow[p \to \infty]{} & -\frac{4 \sqrt{5}}{63} y^2 (1-3 \nu ) \frac{\left(1-e^2\right)^3}{e^2} p^2 \label{eq:NpAsympPlus:42} \, , \\
\frac{N^{4 4}_p}{D_{|p|}^\mathrm{LO}} \xrightarrow[p \to \infty]{} & -\frac{4}{9} \sqrt{\frac{5}{7}} y^2 (1-3 \nu ) \frac{\left(1-e^2\right)^{9/2}}{e^4} p^3 \label{eq:NpAsympPlus:44} \, , 
\end{align}
\end{subequations}

\noindent as well as the leading order $p \to -\infty$ asymptotic expansion of $N^{l m}_p$,

\begin{subequations}
\label{eq:NpAsympNeg}
\begin{align}
\frac{N^{2 0}_p}{D_{|p|}^\mathrm{LO}} \xrightarrow[p \to -\infty]{} & \sqrt{\frac{2}{3}} \left\{1+y^2 \left(\frac{26}{7}-\frac{\nu }{7}\right) \left(1-e^2\right)^{3/2} p\right\}  \label{eq:NpAsympNeg:20} \, , \\
\frac{N^{2 1}_p}{D_{|p|}^\mathrm{LO}} \xrightarrow[p \to -\infty]{} &  \frac{\rmi}{3} y \left(\delta \mu - \frac{3}{2} y \delta\chi \right)  e\label{eq:NpAsympNeg:21} \, , \\
\frac{N^{2 2}_p}{D_{|p|}^\mathrm{LO}} \xrightarrow[p \to -\infty]{} &  e^2 \left\{\frac{5}{4} \frac{1}{\left(1-e^2\right)^{3/2} p}-y^2 \left(\frac{73}{12}-\frac{7 \nu }{4}\right)\right\} \label{eq:NpAsympNeg:22} \, , \\
\frac{N^{3 0}_p}{D_{|p|}^\mathrm{LO}} \xrightarrow[p \to -\infty]{} & \frac{1}{\sqrt{42}} y^2 (1-3 \nu ) \left(1-e^2\right)^{3/2} p  \label{eq:NpAsympNeg:30} \, , \\
\frac{N^{3 1}_p}{D_{|p|}^\mathrm{LO}} \xrightarrow[p \to -\infty]{} &  -\frac{5 \rmi}{12 \sqrt{14}} y \delta \mu  e \label{eq:NpAsympNeg:31} \, , \\
\frac{N^{3 2}_p}{D_{|p|}^\mathrm{LO}} \xrightarrow[p \to -\infty]{} & \frac{5}{24} \sqrt{\frac{5}{7}} y^2 (1-3 \nu ) e^2 \label{eq:NpAsympNeg:32} \, , \\
\frac{N^{3 3}_p}{D_{|p|}^\mathrm{LO}} \xrightarrow[p \to -\infty]{} & -\frac{35}{16}\sqrt{\frac{5}{42}} i y \delta \mu  \frac{e^3}{\left(1-e^2\right)^{3/2} p} \label{eq:NpAsympNeg:33} \, , \\
\frac{N^{4 0}_p}{D_{|p|}^\mathrm{LO}} \xrightarrow[p \to -\infty]{} & \frac{5}{42 \sqrt{2}} y^2 (1-3 \nu ) \left(1-e^2\right)^{3/2} p \label{eq:NpAsympNeg:40} \, , \\
\frac{N^{4 2}_p}{D_{|p|}^\mathrm{LO}} \xrightarrow[p \to -\infty]{} & \frac{\sqrt{5}}{36} y^2 (1-3 \nu ) \left(1-\frac{9 e^2}{4}\right) \frac{e^2}{\left(1-e^2\right)^{3/2} p} \label{eq:NpAsympNeg:42} \, , \\
\frac{N^{4 4}_p}{D_{|p|}^\mathrm{LO}} \xrightarrow[p \to -\infty]{} & -\frac{35}{64} \sqrt{\frac{5}{7}} y^2 (1-3 \nu ) \frac{e^4}{\left(1-e^2\right)^{3/2} p} \label{eq:NpAsympNeg:44} \, .
\end{align}
\end{subequations}

\section{1.5 PN Spin Contributions}
\label{sec:appendix:Spin}

In this appendix, we list the 1.5PN spin corrections to the Fourier mode amplitudes. These corrections affect only the $(2,0)$, $(2,2)$, $(3,0)$, and $(3,2)$ modes. To compute them, we use the 1.5PN expressions for the $K^{lm}$ amplitudes and the quasi-Keplerian parametrization from Ref.~\cite{Henry:2023tka}, following the same procedure outlined in Sec.~\ref{sec:1PN:Fourier}. For the $(2,0)$, $(3,0)$, and $(3,2)$ modes, the 1.5PN spin corrections can be easily incorporated into the 1PN expressions of Eq.~\eqref{eq:Np}, yielding:

\begin{subequations}
\label{eq:NpSpin}
\begin{align}
N^{2 0}_p =& \sqrt{\frac{1}{6}} e \Bigg\{\left[1-y^2 \left(\frac{9}{14}+\frac{17 \nu}{42}+e^2 \left(\frac{19}{14}-\frac{17 \nu }{42}\right)\right) + y^3 \left(\frac{7 \chi_{\text{eff}}}{3}+\frac{\delta \mu  \delta \chi }{3}\right) \right] \mathcal{C}_{1,p}(p e) \nonumber \\ 
& + p y^2 \left(1-e^2\right) \left[\frac{26}{7}-\frac{\nu}{7} -y \left(\frac{3 \chi_\mathrm{eff}}{2}+\frac{\delta \mu  \delta \chi }{2}\right) \right] \mathcal{S}_{1,p} (p e) \Bigg\} \label{eq:NpSpin:20} \, , \\
N^{3 0}_p =& \frac{1}{2 \sqrt{42}} p y^2 \left[1-3 \nu + y \left(\chi_\mathrm{eff}-\delta \mu  \delta \chi \right) \right] \left(1-e^2\right)^{3/2} \mathcal{C}_{0,p}(p e) \label{eq:NpSpin:30} \, , \\
N^{3 2}_p =& \frac{1}{12} \sqrt{\frac{5}{7}} p y^2 \left[1-3 \nu + y \left(\chi_\mathrm{eff}-\delta \mu  \delta \chi \right)\right] \left(1-e^2\right)^{3/2} \Bigg\{\mathcal{C}_{0,p}(p e) \nonumber \\
& - p \left[\sqrt{1-e^2} (\mathcal{C}_{0,p}(p e)-\mathcal{C}_{2,p}(p e))-(2 e \mathcal{S}_{1,p}(p e)-\mathcal{S}_{2,p}(p e))\right]\Bigg\} \label{eq:NpSpin:32} \, , 
\end{align}
\end{subequations}

\noindent where $\chi_\mathrm{eff}$ is the effective inspiral spin parameter~\cite{Damour:2001tu,Racine:2008qv,Ajith:2009bn}, defined as

\begin{equation}
    \chi_\mathrm{eff} = \frac{m_1 \chi_1 + m_2 \chi_2}{m_1 + m_2} \, .
    \label{eq:chi_eff}
\end{equation}

For the $(2,2)$ mode, the expression is more involved. Below we list only the 1.5PN spin correction, which can be added to the 1PN expression in Eq.~\eqref{eq:Np}:

\begin{align}
(N^{2 2}_p)_\mathrm{SO} =& \frac{e}{3} \Bigg\{\left(7 \chi_\mathrm{eff}+\delta \mu  \delta \chi \right) \mathcal{C}_{1,p}(p e)-2 \sqrt{1-e^2} \left(4 \chi_\mathrm{eff}+\delta \mu  \delta \chi \right) \mathcal{S}_{1,p}(p e)\Bigg\} \nonumber \\
& + p \Bigg\{\left(1-e^2\right) \bigg[\sqrt{1-e^2} \left\{\left(\frac{19 \chi_\mathrm{eff}}{2}+\frac{5 \delta \mu  \delta \chi }{6}\right) \mathcal{C}_{0,p}(p e)-\frac{1}{3} \left(\chi_\mathrm{eff}+\delta \mu  \delta \chi \right) \mathcal{C}_{2,p}(p e)\right\}  + \frac{e}{4} \left(3 \chi_\mathrm{eff}+\delta \mu  \delta \chi \right) \mathcal{S}_{1,p}(p e) \nonumber \\
&+\frac{1}{3} \left(\chi_\mathrm{eff}+\delta \mu  \delta \chi \right) \mathcal{S}_{2,p}(p e)\bigg]-\frac{1}{2} \left(7 \chi_\mathrm{eff}+\delta \mu  \delta \chi \right) \left[\left(\sqrt{1-e^2} \mathcal{C}_{0,p}(p e)-e \mathcal{S}_{1,p}(p e)\right)+\frac{1}{3} \tilde{f}_{\beta,p}(e)\right]\Bigg\} \nonumber \\
& + \frac{1}{12} p^2 \left(1-e^2\right)^2 \left(\chi_\mathrm{eff}-\delta \mu  \delta \chi \right) \left\{\mathcal{C}_{0,p}(p e)-\mathcal{C}_{2,p}(p e)+\sqrt{1-e^2} \mathcal{S}_{2,p}(p e)\right\} \, ,
\label{eq:NpSpin22}
\end{align}

\noindent where $\tilde{f}_{\beta,p}(e)$ is defined in Eq.~\eqref{eq:fbetafourier}.

\FloatBarrier
\twocolumngrid

\bibliography{Refs}

\end{document}